\documentclass[twocolumn]{aastex631}
\usepackage[version=4]{mhchem}
\usepackage{xcolor}

\newcommand{\abinit}{\textsl{ab initio}}

\newcommand{\spb}{\sigma^{\text{PB}}}
\newcommand{\YUMI}{\textsc{Yumi}}

\newcommand{\wn}{$\mathrm{cm^{-1}}$}
\newcommand{\mwn}{\mathrm{\,cm^{-1}}}
\usepackage{bbm} 

\usepackage{soul}
\usepackage{color, colortbl}
\definecolor{Gray}{gray}{0.75}
\usepackage{relsize}
\usepackage{subfigure}
\usepackage{colortbl}
\usepackage{enumerate}  
\usepackage{comment}

\newcommand\COTwo{\ce{CO2}}
\newcommand\HTwo{\ce{H2}}

\begin{document}

\title{Comprehensive \textsl{Ab Initio}~ Quantum Computations of \ce{CO2-H2} and \ce{CO2-He} Collisional Properties}

\author[0000-0002-8052-3893]{Prajwal Niraula}
\affiliation{Department of Earth, Atmospheric and Planetary Sciences, MIT, 77 Massachusetts Avenue, Cambridge, MA 02139, USA}
\affiliation{Department of Physics and Kavli Institute for Astrophysics and Space Research, Massachusetts Institute of Technology, Cambridge, MA
02139, USA}
\email{Email: pniraula@mit.edu}

\author[0000-0003-2355-4543]{Laurent Wiesenfeld}
\affiliation{Center for Astrophysics, Harvard \& Smithsonian, Atomic and Molecular Physics Division, 60 Garden Street, Cambridge, MA 02138, USA}
\affiliation{Universit\'e Paris-Saclay, CNRS, Laboratoire Aim\'e-Cotton, 91405 Orsay, France }
\affiliation{Department of Earth, Atmospheric and Planetary Sciences, MIT, 77 Massachusetts Avenue, Cambridge, MA 02139, USA}
\email{Email: laurent.wiesenfeld@universite-paris-saclay.fr;\\ laurent.wiesenfeld@cfa.harvard.edu}

\author[0000-0002-8980-1087]{Nejmeddine Jaïdane}
\affiliation{University of Tunis El Manar, Faculty of Sciences, Tunis, LSAMA, Tunisia}
\affiliation{Universit\'e Paris-Saclay, CNRS, Laboratoire Aim\'e-Cotton, 91405 Orsay, France }

\author[0000-0003-2415-2191]{Julien de Wit}
\affiliation{Department of Earth, Atmospheric and Planetary Sciences, MIT, 77 Massachusetts Avenue, Cambridge, MA 02139, USA}

\author[0000-0002-7691-6926]{Robert J. Hargreaves} 
\affiliation{Center for Astrophysics, Harvard \& Smithsonian, Atomic and Molecular Physics Division, 60 Garden Street, Cambridge, MA 02138, USA}

\author{Jeremy Kepner}
\affiliation{Lincoln Laboratory, MIT, Lexington, MA 02421, USA}

\author{Deborah Woods}
\affiliation{Lincoln Laboratory, MIT, Lexington, MA 02421, USA}

\author{Cooper Loughlin}
\affiliation{Lincoln Laboratory, MIT, Lexington, MA 02421, USA}

\author[0000-0003-4763-2841]{Iouli E. Gordon} 
\affiliation{Center for Astrophysics, Harvard \& Smithsonian, Atomic and Molecular Physics Division, 60 Garden Street, Cambridge, MA 02138, USA}

\begin{abstract}
We present comprehensive \textsl{ab initio} fully quantum calculations of CO$_{\rm 2}$--H$_{\rm 2}$ and CO$_{\rm 2}$--He collisional properties. Our framework combines CCSD(T) potential-energy-surface calculations with close-coupling dynamical scattering in the \YUMI~framework to derive elastic and inelastic cross sections, rate coefficients, and pressure broadening parameters. We characterize the rotational dependence of the broadening coefficients up to $j=25$ for CO$_{\rm 2}$--H$_{\rm 2}$ and $j=40$ for CO$_{\rm 2}$--He, and their temperature dependence over 40--800 K. We also provide Pad\'e fits as a function of rotational quantum number, enabling extrapolation and integration into spectroscopic databases including HITRAN and HITEMP. The resulting pressure broadening coefficients reproduce available experimental measurements on an absolute scale, without empirical correction factors, and meet the $\sim$10\% precision requirement identified for \textit{JWST}-era exoplanet atmospheric studies. This represents a substantial improvement over previously available parameters, which at higher temperatures ($T>400$ K) can fall outside the desired precision by up to a factor of five. All derivations, computed collisional properties, and database-ready products are provided with this manuscript. Together, these results establish a comprehensive \textsl{ab initio}, parameter-free, fully quantum foundation for CO$_2$ collisional broadening by H$_2$ and He, while demonstrating the transformative potential of the ab-initio approach for next-generation spectroscopic needs across planetary atmospheres, combustion, health sciences, and fusion-plasma diagnostics.
\end{abstract}



\section{Introduction} 
\label{sec:intro}

\COTwo-\HTwo \ and \COTwo-He \ collisional systems have a wide range of applications. As a major carbon-bearing molecule and a strong infrared absorber, \COTwo~has now been detected with JWST in an increasing number of exoplanet atmospheres spanning a broad range of sizes and temperatures \citep{fu2025}. Constraining the amount of \COTwo~is a powerful diagnostic to inform the formation and evolution history of an exoplanet \citep{oberg2011}, a task that hinges on accurate determination of its collisional properties. Similarly, global circulation models governing the climates of these exoplanets are sensitive to the collisional properties of \COTwo~\citep{chaverot2023}. For gas and ice giants in the solar system, \COTwo~is not abundant but remains an important tracer for atmospheric processes and external material delivery \citep{feuchtgruber1997}.  Similarly, \COTwo~is often found in abundance in protoplanetary disks  \citep{bosman2017, Grant2023, Frediani2025}. Naturally, this system is of interest beyond astrophysics, lying at the center of remote-sensing applications for fields ranging from combustion and the petrochemical industry to medicine.  

As observational data improve, increasingly precise scientific inferences require correspondingly accurate opacity models. In the context of exoplanetary sciences, \citet{Niraula2022} showed that current limitations in line-shape parameters, including broadening coefficients and far-wing behavior, impose significant accuracy bottlenecks on our ability to characterize exoplanetary atmospheres. Unfortunately, broadening parameters for many relevant collisional systems are still lacking \citep{tan2022}. Experimental avenues require substantial funding, a large workforce, and time commitments \citep{fortney2016}. Moreover, some of the experiments, including \COTwo-\HTwo~at elevated temperatures, are non-trivial or simply dangerous.  To address this gap, we demonstrated an alternative \abinit\, first principle-based computational approach in \citet{Wiesenfeld:2025aa} targeting a single transition of \COTwo-\HTwo, the only transition with experimental values at different temperatures \citep{hanson2014} available at the time. 
 
Here, we expand upon \citet{Wiesenfeld:2025aa} and perform a comprehensive calculation for the collisional properties of \COTwo-\HTwo~and \COTwo-He. The latter system has been substantially better studied than \COTwo-\HTwo\, both experimentally and theoretically (see, for instance, \citet{Thibault:2000aa, Korona2001}), but is still important to benchmark our calculations, particularly due to its renewed interest by its relevance to exoplanetary atmospheres \citep{Hendaoui2025, HendaouiICARUS}. In \citet{Wiesenfeld:2025aa}, we determined the precision requirement to power \textit{JWST} exoplanet studies as a $\leq$10\% precision on pressure broadening coefficients. Each step in our framework is carefully designed to support this goal. With this framework, we compute ro-vibrational transitions up to $|m|=40$ (that is $j'$=40, $~j''$=41) (see \autoref{sec:results}), and temperatures spanning 40--800~K, and derive the pressure broadening dependence on rotational quantum number and temperature. 

Beyond producing a new dataset for \COTwo-\HTwo~and \COTwo-He, this work serves as a validation of a scalable first-principles opacity-generation framework. A key test is whether a fully \abinit\ calculation can reproduce available measurements on an absolute scale, without empirical correction factors, while producing database-ready outputs across the temperature and rotational ranges required by modern remote-sensing applications. Scaling this approach to more complex systems also requires traceable, AI-assisted, expert-validated computational workflows for deriving, implementing, and validating increasingly complex coupled-channel formalisms.

We introduce key definitions, the state of the art, and our framework in \autoref{sec:framework}, a summarized theory of collisional properties in \autoref{sec:pbps_theory}, and our methods in \autoref{sec:theo}. We present our results in \autoref{sec:results} and compare them to experimental values in \autoref{sec:compare}. Concluding remarks are offered in \autoref{sec:conclusion}.

\section{Definitions, State of the Art, \& Framework}\label{sec:framework}

\begin{figure}[t!]
    \includegraphics[width=\linewidth,trim = 3.25cm 0.1cm 1.75cm 0.5cm, clip=true]{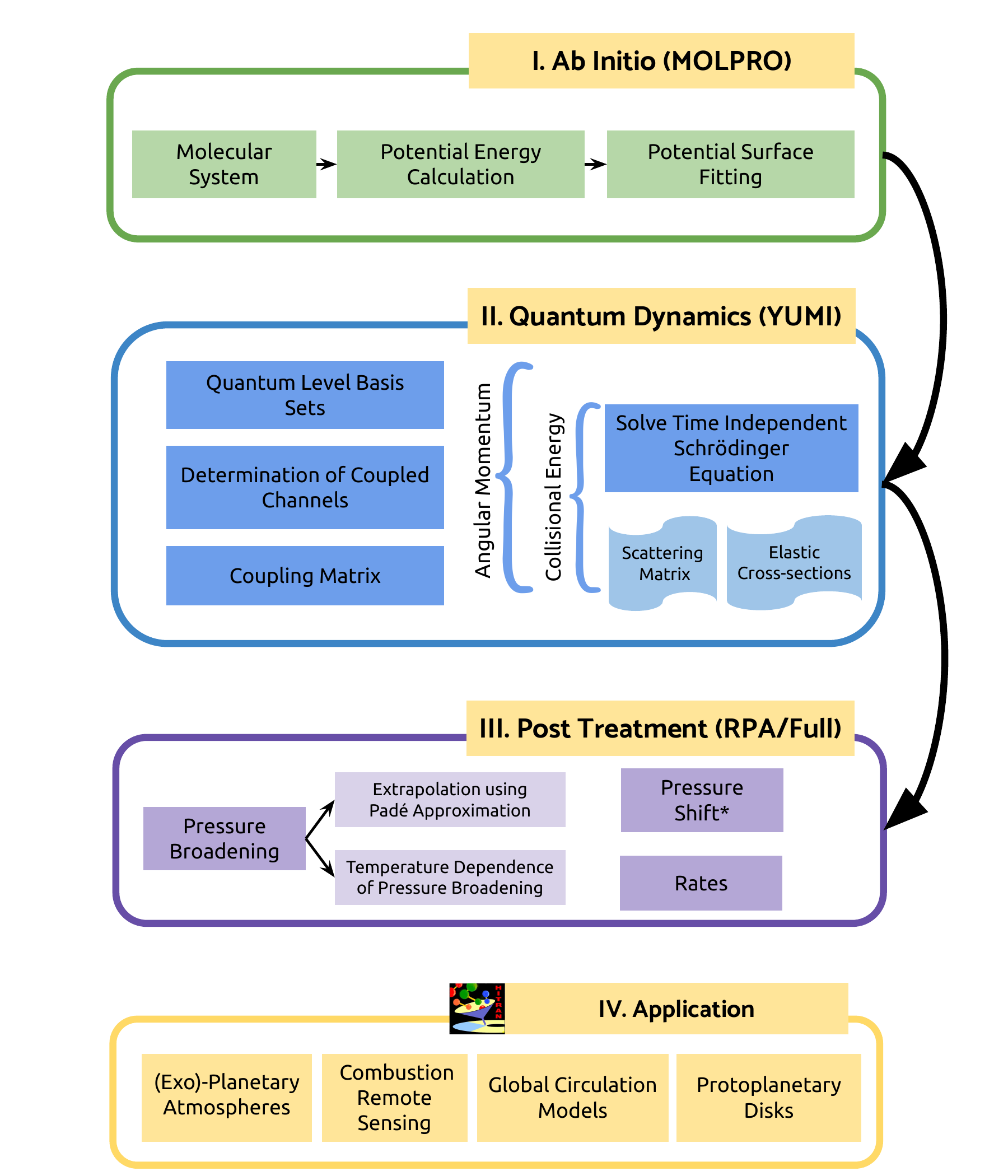} 
    \caption{ Schematics of the deployed framework for calculating collisional properties of \ce{CO2}-He and \ce{CO2}-\HTwo. Various collisional properties of \COTwo~by \HTwo~ and He are estimated, which is useful for a wide range of applications, including exoplanetary retrievals. {*}The calculation of pressure shift is deferred to follow-up work.}
    \label{fig:Introduction}
\end{figure}

\begin{figure}[ht!]
    \includegraphics[width=\linewidth,trim = 0.25cm 0.1cm 0.0cm 0.1cm, clip=true]{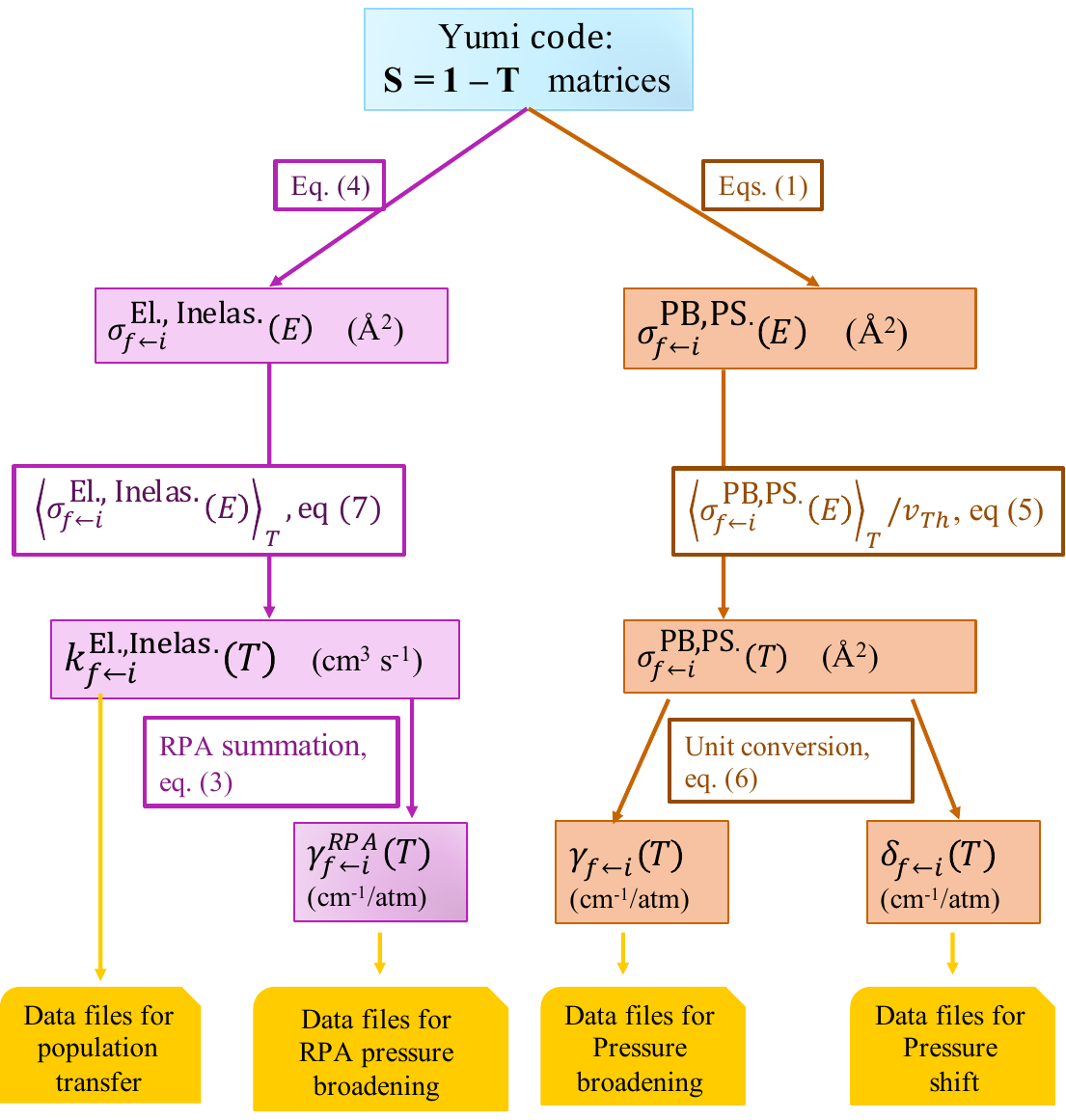}
    \caption{ Details of the computational flow for the post-treatment following the dynamics calculation with \YUMI.}
    \label{fig:flowcomput}
\end{figure}

By \abinit, we mean here that the only inputs of our calculations are molecular structures and constants: the average geometry of \ce{CO2} and \ce{H2} molecules in their relevant vibrational levels and their corresponding rotational constants. All other quantities are computed, including electronic structures of the \ce{CO2-H2} and \ce{CO_{2}-He} van der Waals complexes (see \autoref{sec:abinitio}), and all dynamical quantities (elastic and inelastic cross sections, transfer of population rates, pressure broadening rates, see \autoref{sec:results}). An overview of the framework is presented in \autoref{fig:Introduction} and the computational flow in \autoref{fig:flowcomput}. \YUMI\, is our primary in-house code for performing dynamical calculations, which will be described in detail in an upcoming paper (Jaïdane et al., in prep.).

A central challenge in developing this framework was not only computational cost, but also the reliable translation of published pressure broadening formalisms into a consistent numerical implementation. The relevant literature spans multiple equivalent but notationally distinct formulations, with different angular-momentum coupling schemes, phase conventions, normalization choices, and degeneracy factors. These differences were not merely cosmetic: in earlier versions of the implementation, they led to discrepancies that had to be traced across published derivations, existing scattering codes, and our own post-treatment routines. AI-assisted tools played an important role in this process by helping compare published formalisms, identify notation and normalization inconsistencies, support code review and debugging, and organize large-scale computational outputs. All physical assumptions, equations, numerical implementations, and results were evaluated and validated by the authors. In particular, special care was taken to validate our computational chain against well-established results of the \ce{CO - H2} collisional system, for which a thoroughly validated energy is available \citep{wernli2006improved,thibault_line_2025}.

This framework may be compared to the commonly used approach in the literature for the \abinit\, computation of inelastic scattering. Inelastic cross sections are used for the interpretation of spectral intensities and properties of interstellar matter, especially for rotational spectra \citep{Goldsmith:1999aa}. Numerous studies have used  \abinit\, methodologies 
\citep[for example see:][]{atoms8020015,1960RSPSA.256..540A,valiron2008,2013A&A...553A..50D,atoms8020015,Dubernet:2005aa,Sahnoun2018,WOS:000553704600007,Bergeat:2020aa,2012PhRvA..86b2705D,WOS:000951897400001,2023CoPhC.28908761A,Selim:2023aa}.  This large body of literature uses the same type of \abinit\ approach, though their specific goals differ widely from our current focus, in terms of precision, temperature range and type of cross-sections (see \autoref{sec:dynamics}). Inelastic cross-sections and transfer of population rates are necessary for evaluating spectral line intensities for fairly common conditions in the interstellar media (critical density $n^*(\mathrm{H_2)\sim10^{\,4-6}\,cm^{-3}}$, temperatures T$\sim$5-300 K). However, the precision needed for interstellar studies is modest, as even a 50\% error on the rates remains acceptable. This contrasts sharply with the stringent precision required here, i.e 10\% \citep{Wiesenfeld:2025aa}.

In addition, some pressure broadening computations have been computed \abinit\, akin to ours, with publicly available codes (like \textsc{MOLSCAT} -- \cite{hutson_molscat_2019}). In particular, Thibault et al. computed several pressure broadening of molecules by either He or rare gases, \citep[see][]{Thibault:2000aa,Thibault:2001aa,WOS:000311303400013,WOS:001152493800001}. Most of these calculations compare favorably with the experimental results, especially so in the intermediate temperature ranges ($T\sim 300$~K). The group from Torun has been carrying out fully ab initio calculations, with home-grown scattering code,  targeting a very high level of precision to obtain non-Voigt parameters (e.g. \cite{WOS:000553704600007,2023JChPh.159m4301O,2024A&A...687A..69J}) for gases of atmospheric interest, but the targeted sub-percent precision allows calculation of only a few transitions at a time. We computed pressure broadening (and shifts) for \ce{H2O} in collision with \ce{H2} at low temperatures ($T\lesssim  100 $~K), and compared those with experimental results. Despite experimental difficulty, excellent agreement was found between them and  \abinit\, theory \citep{2012PhRvA..86b2705D}.

Several approaches have been used to model pressure broadening, including approximate methods that can handle dynamically intricate cases. A comprehensive review, including different formalisms, has been given by \cite{hartmann2021}. Classical models, in particular, have been used for many years to provide order-of-magnitude estimates of pressure broadening, and in some instances, where empirical corrections are possible, reliable predictions are possible such as for C$_2$H$_2$ broadened by \HTwo\,\citep{2025JQSRT.33009225S}. Classical (requantized) molecular dynamics simulations have also been successful (see, for instance, \cite{2025JQSRT.33109264N}).

\section{On Pressure Broadening and Shift}\label{sec:pbps_theory}
We present here a condensed theory of pressure broadening and pressure shift (PB/PS) coefficients, including the relevant approximations. Although the theory has long been established, multiple formulations of the equations exist, and we therefore find it useful to clarify the specific approach we adopt here. Early quantum derivations, still in use, can be found in \cite{1958PhRv..112..855B,ben-reuven_impact_1966,GREEN1977119,Schaefer:1987aa}. A comprehensive analysis together with different approximations, as well as several extensions of the theory, can be found in \citet{hartmann2018}.

Here we make explicit use of the  \textit{(i)} impact approximation, \textit{(ii)} the isolated line approximation, and \textit{(iii)} isolated events approximation, \citep{hartmann2018}. The impact approximation supposes that the duration of the collision is so short that no dynamics of both target and projectile occur \emph{within} the time of the existence of the complex. Isolated line approximation neglects the effects of line mixing.  Although carbon dioxide is notorious for having strong line mixing in the Q-branches of the ``perpendicular'' bands, including the bending fundamental, these are narrow features and not present in the asymmetric stretch fundamental, which has been identified on exoplanets so far. Moreover, there are ways to approximately estimate first-order line mixing using energy power gap (EPG) approximation, as it was done for air and self-broadened values in \citet{10.1016/j.jqsrt.2020.107283}.  Isolated events allow for collisions to always be two-body events, in distinct succession; no three-body events are taken into account. This latter approximation holds, in practice, up to approximately 1 amagat density for the gas (number density of one standard atmosphere,  at 0 $^{\circ}$C; it amounts to $\approx 2.687 \times10^{19} \text{cm}^{-1}$).

Let $\sigma^{\text{PB}}(E)$ be the pressure broadening cross section as a function of $E$, the collision energy, and let $\sigma^{\text{PB}}(T)$ be the average of $\sigma^{\text{PB}}(E)$ over the Maxwellian distribution of the collision energy $E$ (energies are defined in the Appendix, section \autoref{sec:notations}). A complete derivation of the pressure broadening (PB) and pressure shift (PS) cross sections is found in the seminal papers \cite{ben-reuven_impact_1966,coombe_definitions_1975}.

The full case (with $j_2\geq$0) is presented in the following equation, defining micro-canonical (energy-dependent) cross-sections :

\begin{widetext}
\begin{eqnarray}\label{eq:section}
    \sigma^{\text{PB;PS}}_{j_1''\leftarrow j_1'}(E) &= &\,\frac{\pi}{k^2} N\left(j_2\right)
    \sum_{j'_{12},j''_{12},\overline{j'_{12}},\overline{j''_{12}}}\;
    \sum_{\ell,\overline{\ell},\overline{j_2};J',J'' }\;
    X\left(j'_1,j''_1;j_2,\overline{j_2};\ell,\overline{\ell}; j'_{12},j''_{12}, \overline{j'_{12}},\overline{j''_{12}}; J',J''\right) \times \\ \nonumber
      &   & \left[ \delta_{\ell,\overline{\ell}}\,\delta_{j_2,\overline{j_2}}\,\delta_{j''_{12},\overline{j_{12}''}}\, \delta_{j'_{12},\overline{j'_{12}}} \quad - \;
      \left<\overline{\ell}, \overline{j''_{12}} , j_1'', \overline{j_2}|\mathsf{S}^{J''}(E)|\ell, j_{12}'',j_1'',j_2\right>^{*}
  \left<\overline{\ell}, \overline{j'_{12}}, j_1',\overline{j_2}|\mathsf{S}(E)^{J'}|\ell j'_{12}j_1'j_2\right>\right].
\end{eqnarray}
\end{widetext}

Pressure broadening and shift are respectively the real and imaginary parts of the complex number $ \sigma^{\text{PB;PS}}_{j_1''\leftarrow j_1'}(E)$. $k=\sqrt{2\mu E /} \hbar$ is the wavenumber of the complex at collision energy $E$, reduced mass $\mu$. $j_1'$ and $j_1''$ denote respectively the initial and final rotational angular momentum of \ce{CO2}. $\mathsf{S}(E)= \mathbf{1} - \mathsf{T}(E)$ is the scattering matrix, with $^*$ denoting complex conjugation.   $J',\,J''$ are the total angular momenta of the collision.  Note that both quantum numbers $\ell$ (orbital angular momentum) and $j_2$, (projectile rotational quantum number) are the same for both brackets. $N=1/(2j_2+1)$ is the normalization, which takes into account the degeneracies of the incoming projectile. 

The important points to observe in \autoref{eq:section} are: \textit{(i)} both brackets in the equation are elastic transitions, that is, the internal states of the observed molecule \ce{CO2} do \emph{not} change in the interaction, \textit{(ii)} environment quantum numbers need not be elastic, but are the same for both brackets, and \textit{(iii)} both interactions occur at the same kinetic energy but at different total energies $\mathcal{E}_{\text{tot}}=E+E_{\text{internal}}$.

The $X(\cdot)$ coefficients are derived by integrating the differential amplitude of scattering over the spherical angles and using the form of the rotational eigenfunctions. They \emph{do not depend} on the vibrational quantum states, thereby justifying our neglect of the vibrational dynamics, except for the symmetry associated with them, see \autoref{sec:spy}. Full formulae can be found in \cite{Schaefer:1987aa,coombe_definitions_1975}, even if vibrational quantum numbers are not explicitly displayed.  We have the quantum numbers: $(j_2,\ell,j_{12})$, where $\ell$ is the orbital quantum number (the angular number of the projectile with respect to the target center-of-mass) and $j_{12}$ is the recoupling of $j_1$ and $j_2$, see section \autoref{sec:notations} for detailed descriptions.  We use this form, which we verified to be exactly equivalent to other definitions, thanks to the invariance properties of 6-$j$ symbols:

\begin{widetext}
\begin{eqnarray}
\label{eq:X}
  X(\cdot)&=& P\left(\ell+\bar{\ell} +j''_{12}-\overline{j''_{12}} + j'_{12}-\overline{j'_{12}} +j'_1-j''_1+ j_2 - \overline{j'_2}\right)\left[J' J''\right]
    \left[     j'_{12} \,j''_{12} \, \overline{j'_{12}} \, \overline{j''_{12}}\right]^{1/2} \times \nonumber\\ 
   & &\begin{Bmatrix} J''& J' &q\\ j'_{12}&  j''_{12}& \ell \end{Bmatrix}\,
      \begin{Bmatrix} J''& J' &q \\ \overline{j'_{12}} & \overline{j''_{12}} & \overline{\ell}\end{Bmatrix}\,
      \begin{Bmatrix} j_1'' & j_1' & q \\  j'_{12}&  j''_{12}& j_2 \end{Bmatrix}\,
      \begin{Bmatrix} j_1'' &j_1' &q \\ \overline{j'_{12}} &  \overline{j''_{12}}& \overline{j_2} \end{Bmatrix},
\end{eqnarray}
\end{widetext}
where $P(e)= (-1)^e$ and $[JJ']=(2J+1)(2J'+1)$. $q$ is the multipolar order of the electromagnetic transition. Here $q=1$, for dipolar transition. The normalization and angular recoupling schemes (Eq. \eqref{eq:section}, Eq. \eqref{eq:X}) are valid for any collision rotator - atom or rotator - rod, as the supplementary quantum numbers $k$ or $\tau$ for resp. symmetric or asymmetric rotors 
are spectators in the recoupling scheme presented in \autoref{sec:notations}.  However, for the rotor/rotor  (e.g., like water-water collisions) $X(.)$  coefficients are different, as the projectile rotational eigenfunctions in the lab. frame will entail full Wigner rotation functions \citep[see e.g.][]
{van_der_avoird_intermolecular_1994}. Noteworthy is the appearance of the $g(q')$ denominator ($g=1/\left(2j'_2+1\right)$ for a rod), as the degeneracy of initial conditions in the spectator projectile must be taken into account, see \citep{GREEN1977119,2012PhRvA..86b2705D}. One must also ensure that the  $\sigma^{\text{PB;PS}}_{j_1''\leftarrow j_1'}(E)$ sections are independent of the order $i\leftrightarrow f$, and so, adapt the degeneracies included (or not) in the $\mathsf{T}$ matrices computations, \autoref{sec:norm_cross}. 

Also, in the case of helium or para-\ce{H2}, $j_2=0$ (no structure of the projectile), \autoref{eq:section}, \autoref{eq:X} simplifies thanks to the properties of the Wigner 6-j symbols, yielding an equivalent expression to Eq. 3 in \citet{Thibault:2000aa}.

In addition to the above formulae, the use of the optical theorem allows us to have another view of the PB cross-sections, which is fully equivalent  (but not valid for the PS cross-sections) \citep{1958PhRv..112..855B, Faure:2013aa}:

\begin{widetext}
\begin{equation}
  \label{eq:RPA}
    \spb_{j''_1\leftarrow j'_1} (E)  = 
    \frac12\,\left[\sum_{\bar{j_1}\neq j'_1 }\sigma^{\text{In.}}_{\bar{j_1}\leftarrow j'_1}(E) 
    +  \sum_{\bar{j_1}\neq j''_1 }\sigma^{\text{In.}}_{\bar{j_1}\leftarrow j''_1}(E) \right]  
    +\int\left| f_{j'_1}(E,\Omega)-f_{j''_1}(E,\Omega) \right|^2 \mathrm{d}\Omega\quad ,
\end{equation}
\end{widetext}
where $ \sigma^{\text{In.}}_{\bar{j_1}\leftarrow j'_1}(E)$ are ordinary inelastic cross sections, $f_{j'_1}(E,\Omega)$ are elastic differential scattering amplitude, and $\Omega$ is the solid angle of scattering. The second term, which describes interferences between incoming and outgoing scattering wavefunctions, may be neglected (the Random Phase Approximation (RPA)) at higher energies, when forward scattering dominates the dynamics and phase differences oscillate rapidly, therefore goes to zero when averaged over a Maxwellian distribution of kinetic energies. We pursue both approaches in this work, Eq.\eqref{eq:section} and Eq.\eqref{eq:RPA}, and test their validity.

In order to be complete, the ordinary cross sections, elastic or inelastic, are defined as:
\begin{equation}\label{eq:secord}
    \sigma_{f\leftarrow i}(E) = \frac{\pi}{k^2} \frac{1}{g_i} \sum_{qq'} \left(2J+1\right)\left|\left <fq'|\mathsf{T}^J|iq\right >\right|^2
\end{equation}
where $i$ and $f$ are quantum numbers of initial and final levels (in all generality), $q$, $q'$ are the other quantum numbers describing the couplings of the various states, $g_i$ is the degeneracy of the initial level depending on the type of section that we are interested in (see Eq.(12) in \citet{GREEN:1975ae}), and $k$ is as in Eq.\eqref{eq:section}.

The temperature-dependent average section (in \AA$^2$) and the pressure broadening coefficient  $\gamma(T)$ (in wavenumber/atmosphere, for Half Width at Half Maximum) are defined as:
\begin{align}\label{eq:gamma}
    \sigma^{\text{PB;PS}}_{j''_1\leftarrow j'_1} (T) & = k^{\text{PB;PS}}_{j''_1\leftarrow j'_1}\,/\,v_{\text{Th}} \\
    \gamma & = k^{\text{PB}}_{j''_1\leftarrow j'_1}/2\pi k_BT=\spb(T)\;v_{\text{Th}}\,/\,2\pi k_BT
\end{align}
where $k_{j_1''\leftarrow j'_1}(T)$ is the usual rate coefficient (in cm$^3$ s$^{-1}$) obtained by averaging $ \sigma_{j''_1\leftarrow j'_1} (E) $  over the Maxwellian distribution of $E$ and $v_{\text{Th}}=\sqrt{8k_BT/\pi\mu}$ ($\mu$, reduced mass of the collision, $k_B$, Boltzmann constant):
\begin{equation}\label{eq:rate}
k(T) = \frac{v_{\text{Th}} }{(k_BT)^2}\int_0^\infty   \sigma(E) E \exp(-E/k_BT) \,\mathrm{d}E
\end{equation}
where cross-sections $\sigma(E)$ stand for any type of collisional rate (elastic, inelastic, pressure broadening, pressure shift).

\section{Methods} \label{sec:theo}

\subsection{Spectroscopy}
\label{sec:spy}
Although very well known for many years, both the $^{12}$C$^{16}$O$_2$ and $^1$H$_2$ symmetries are relevant to the present work and are discussed here.  Both molecules show $D_{\infty h}$ symmetry. $^{16}$O has nuclear spin $I=0$ and $^{1}$H has nuclear spin $I=1/2$. Both come in two spin modifications. CO$_2$, with zero nuclear spin, exists for the ground vibrational state only in the para state (angular momentum $j_1=0,2,\ldots$). For asymmetric vibrational states, such as $v_3=1$ (asymmetric stretch),  only odd values of $j_1$ exist. For H$_2$, the two spin modifications exist regardless of the vibrational state.  Para states (singlet total nuclear spin state $I=0$) have even $j_2$ angular momentum, ortho states (triplet total nuclear spin states, $I=1$) have odd angular momentum $j_2$. 

We suppose that the vibrational dependence of pressure broadening is minimal for \ce{CO2}, as is experimentally demonstrated for collisions of \ce{CO2} with air \citep{10.1016/j.jqsrt.2020.107283}. Since the $\nu_3$ band transitions connect levels with $\Delta j_1=\pm 1$ ($P$ and $R$ branches), we need to deal with two different vibrational levels of opposite parity. We considered the vibrational ground state, $v=0$, and the first asymmetric stretch level, $v_3=1$. We used the same PES for both levels, neglecting the slight difference that could arise. 

To summarize, the computations follow these steps: \textit{(i)} select a set of collision energies $E_i$; \textit{(ii)} For each $E_i$, compute for each value of $m$ (see \autoref{sec:notations} for a definition of $m$) the total energy (for example, $m=-25 \Rightarrow j'_1=25\; \text{and} \; j''_1=24$ so that $E'_{tot}= E_i+E_{rot}(25) \;\text{and}\; E''_{tot}= E_i+E_{rot}(24)$; \textit{(iii)} conduct dynamics at $E'_{tot} \Rightarrow \mathsf T'(E_i)$ and  $E''_{tot} \Rightarrow \mathsf T''(E_i)$. Combine via Eq. \eqref{eq:section} to get $\sigma^{\text{PB,PS}}(E_i)$.

\subsection{\textsl{Ab initio} quantum chemistry}
\label{sec:abinitio}
We compute the interaction of \ce{CO2} with the bath species He and \ce{H2}. The following physical parameters (distance in bohr) were used for bond lengths: $R(\mathrm{^{12}C^{16}O}$) = 2.1944, $R({\ce{^1H^1H}} )= 1.448736 $. The \COTwo~average distance depends very weakly on the vibrational state, so that we did not compute a different PES for \ce{CO2}, in the ground or excited vibrational state. The distance HH corresponds to the average ground state of the \ce{H2} molecule \citep{valiron2008}.

The \ce{CO2-He} Potential Energy Surface (PES)  has been computed in several instances with very similar results. Most PES were computed with both spectroscopy of the complex and scattering in mind \citep{Yang:2009aa,2022JChPh.156j4303G}. The \ce{CO2-H2} PES has been computed only in a few instances \citep{Li:2010aa}. The more recent work by \citet{hellman2025} computed potentials taking into account higher order effects, however these effects are not expected to be relevant for our calculations. We, therefore, recompute both PES in an identical fashion. 

We employed the so-called Golden Approximation to compute the Born-Oppenheimer PES \citep{2019JChPh.151g0901K}. At the level of precision required -- about one wavenumber at the bottom of the PES well-- the CCSD(T) method was adopted. This approach was selected because of its versatility, extensive use in previous studies, and balanced trade-off between computational cost and precision \citep{jeziorska2008,varandas_cbs_2018}. In particular, we adopted the option of using the CCSD(T) functional, without the F12 approximation to the electron-electron short-range interaction, because of well-known problems at large intermolecular distances. Instead, we took the slightly more expensive approach of using both mid-bond atomic basis sets \citep{Shaw:2018aa} and a standard Complete Basis Set extrapolation scheme \citep{2019JChPh.151g0901K}. The basis sets used are aug-pVXZ \citep{MOLPRO_brief}, most appropriate for molecules comprising light elements with long-range interactions. We used X values of 3, 4 and, for a reduced set of configurations, X = 5 and 6. We checked that the 3-4 and 4-5 extrapolation schemes are equivalent at the level of precision necessary here. The mid-bond point was situated at half distance between the center of mass of the projectile and the nearest atom of the target, thereby avoiding mid-bond functions that could strongly overlap with target ones.

For computing the interaction energy and building the PES, 
we used the super-molecular approach, corrected for the Basis-Set-Superposition-Error (\citet{2019JChPh.151g0901K}, see \citet{Wiesenfeld:2025aa} for a method description, and \citet{jeziorska2008,varandas_cbs_2018} for the CBS/BSSE formalism). We have:
\begin{equation}
    E^{\text{Int.}}(AB)= E_{AB}(AB)-E_{AB}(A) - E_{AB}(B)
\end{equation}
where $A$ and $B$ denote atomic bases centered on fragments A and B, respectively. The influence of the compound electronic basis set $AB$ is thus taken into account in the same way when computing the electronic energies of the A fragment, the B fragment, and both A and B fragments interacting. 

A few effects are neglected in this approach, that could be of importance as the collision energy increases, such as: \textit{(i)} the rigid body approximation, that neglects the deformation of fragments as they approach one another; \textit{(ii)} neglect of any non Born-Oppenheimer effects, such as possible electronic excited levels contamination of the ground electronic level considered here. Point \textit{(i)} is in line with the impact approximation for the PB formalism. Since all elements are light, no relativistic models of the atoms are necessary, as admitted for elements of the first lines (up to Ne, and even Ar).

The \ce{CO2-He} PES potential $V(r,\theta)$ depends on one angle (the polar angle $0\leq \theta < \pi$)  and on the $r$ distance between atoms C and  He. It makes the computation very economical. We resorted to $n_{\theta}=41$ angles (randomly distributed over $0\leq \theta \leq \pi$), not enforcing the symmetry to have a more robust fit. The random distribution has the advantages of \textit{(i)} being resonance free for all orders of Legendre polynomials $P_l(\cos\theta)$, as long as $l<41/2$, and \textit{(ii)} being able to add points if judged necessary to ensure a good fit. We computed the potential $V(r,\theta)$ for 43 distances, $3.50 \leq r \leq 50$, with steps representative of the variation of $V$. In total, $N\lesssim 2000$ points were computed. The resulting PES is compared to earlier computation from \citet{2022JChPh.156j4303G}.

The \ce{CO2-H2} PES depends on three angles (Two polar angles $0\leq \theta_{1,2} < \pi$ and one dihedral angle $0\leq \phi \leq \pi/2$). We also resorted to a random distribution of the 3 angles, to which we added the special orientations $\left(\theta_1,\theta_2,\phi\right)=(0, 0, 0;\; 0, 90, 0;\;0, 90, 90) $ degrees. Please see \citet{Wiesenfeld:2025aa} for a more detailed discussion on the PES.

\subsection{Fit}\label{sec:theofit}

Both PES were fit on the usual base suitable for scattering computations \citep{valiron2008, hutson_molscat_2019}. The formalism is essentially the same for any scattering involving a target and an atom or a rod. The usual formalism follows \cite{GREEN1977119}:
\begin{equation}
    V(r,\Omega)= \sum_{q}V_q(r)\,A_q(\Omega)
\end{equation}
where $\Omega$ represents the angles setting the orientation of the projectile (here He or \ce{H2}) with respect to the target (\ce{CO2}), and $q$ refers to the appropriate orders of the fitting functions.

For a \ce{H2} projectile, we have the following expression \cite{GREEN1977119}, compatible to our definitions of Legendre $P_l(\cos\theta)$ and associated Legendre functions $\mathcal P_{\ell}^m (\cos\theta)$ :

{
\begin{widetext}
\begin{eqnarray}\label{eq:fit1}
A_{\ell\, \ell_1 \, \ell_2}(\theta_1,\theta_2,\phi)& = & \frac{\left(2\ell+1\right)}{4}(\pi)^{-3/2}\left[\left(-1\right)^{\ell_1-\ell_2}
\begin{pmatrix}
\ell_1 && \ell_2 && \ell \\
0      && 0      && 0
\end{pmatrix}
\mathcal P^0_{\ell_1}(\cos \theta_1) \mathcal P^0_{\ell_2}(\cos \theta_2)\right. \nonumber \\
& & +\left. \sum_{m>0}2 \left(-1\right)^{m+\ell_1-\ell_2}
\begin{pmatrix}
\ell_1 & \ell_2 & \ell \\
m      & -m     & 0
\end{pmatrix}
\mathcal P_{\ell_1}^m(\cos\theta_1)
\mathcal P_{\ell_2}^m(\cos\theta_2) \cos(m\phi)\right]
\end{eqnarray}
\end{widetext}
}

with angles as in \autoref{fig:geometry}. $\big(\begin{smallmatrix}
  . & . & .\\
  . & . & .
\end{smallmatrix}\big)$ is a 3-$j$  symbol. Eq. \eqref{eq:fit1} simplifies if the projectile is an atom ($ m\equiv 0,\, \ell_2\equiv 0,  \ell_1=\ell$) and we have:
\begin{equation}\label{eq:fit2}
A_{\ell_1}(\theta) = P_{\ell_1}(\cos\theta), \,\, \ell_1=0,2,\ldots
\end{equation}
with $P_{\ell_1}(\cos\theta)$, the Legendre polynomials. Non equivalent normalizations of the Legendre polynomials $P_\ell(\cos \theta)$ and the associated Legendre functions $\mathcal P_{\ell}^m(\cos\theta )$ are described in  \autoref{sec:norm_legendre}. They are duly used in the \YUMI\,code.

\subsection{Dynamics}\label{sec:theodyn}

We solve the time-independent Schrödinger equation in the Space-Fixed (SF) reference frame. The Potential Energy Surface (PES) described in \autoref{sec:abinitio} and fitted as described in \autoref{sec:fits}, $V(r,\Omega)$, is the PES introduced in the Schrödinger equation describing the motion of the projectile with respect to the target.

The equations to solve are the Close-Coupling equations (CC) (named Coupled Channels in some literature), as described by \cite{1960RSPSA.256..540A,GREEN1977119}. These equations are coded into the \YUMI\, code, with distinct implementations corresponding to the various geometrical cases (similar to the \textsc{molscat} code, but with a fully modular construction).

With the definition of the expansion as in \autoref{eq:fit1}, the matrix elements of the potential coupling between the states is an algebraic closed formula (Eq.(9) from \cite{GREEN:1975ae}), obtained by integrating the $A_{\ell_1\, \ell_2 \, \ell}(\theta_1,\theta_2,\phi)$ functions over the angles of both rotator eigenfunctions.

The main numerical task is thus to integrate the Schr\"odinger equation in the radial coordinates (the CC equations). In order to allow for the stability of the numerical procedure, radial equations are solved by means of a Riccati equation, for the log-derivative of the $Y_q(r)=\psi'_q(r)/\psi_q(r)$ functions.
The \textsf{T} or \textsf{S}  matrices are determined by matching the relevant $Y_q$ to the spherical regular or irregular Bessel functions, in the asymptotic region \citep{johnson:1973aa,manolopoulos_improved_1986}.

The strategy is to compute the \textsf{T}-matrices for as few kinetic energies as possible, but still enough to have a good description of the scattering and with sufficiently high $E$ to describe the higher temperatures, high $j_1$ cases with reasonable precision. Following \cite{Wiesenfeld:2025aa}, we target convergence at a 10\%-precision level. 

We optimized the computational procedure and the energy/angular parameters (see \autoref{sec:scalability}). For the higher $E$ cases ($E\lesssim 2000\, \mathrm{cm^{-1}}$), the convergence with total angular momentum $J$ is very slow and a value of $J\sim 120$ is necessary, putting a stress on the computation of the Wigner 6-j coefficients. Similarly, in order to treat the couplings properly, enough \ce{CO2} rotational levels must be included. Closed channels with  $E=E_{tot}-E_{rot} > -V_{min}$ must be included; This amounts in practice to adding 3 to 5 closed channels to the basis set. 

While for collisions with He and with para-\ce{H2} ($j_2=0$) these conditions could be met, convergence for ortho-\ce{H2} ($j_2=1$) could not be achieved at $\mathcal{E}_{\text{tot}} \gtrsim 2600\, \mathrm{cm^{-1}}$, because both of the length of the computation and the imprecision of the Wigner 6-j and 9-j coefficients at arguments $\gtrsim 150$. Indeed, the triple degeneracy of the $j_{12}=j_1,\, j_1\pm 1$ leads to a ninefold increase in the computational time (see \autoref{sec:scalability}). As $j_1$ grows, the computational load becomes increasingly prohibitive, making exact CC calculations difficult. For practical purposes, computations were limited to $j_1\leq 24/25$ for collisions with H$_2$ and $j_1\leq 40/41$  for collisions with He. We defer a full analysis for the $j'_2>1$ cases to a follow-up work. Studies to mitigate these problems are underway, either by using more robust code or by allowing for relevant approximations.

\subsection{Post-Treatment}
\YUMI~  yields  tables of the $\left<|\mathsf{T}^J| \right>$ complex matrix elements. These are duly summed to yield the observables needed as shown in Eq.\eqref{eq:section},  or Eq.\eqref{eq:secord}. The same procedure had been used by us for differential cross sections \citep{Yang:2011aa}, or previous pressure broadening sections \citep{2012PhRvA..86b2705D}. The \textsc{Molscat} code, used in \citet{2012PhRvA..86b2705D}, follows a slightly different approach, but is fully equivalent for computing the observables, even if considerably slower.

\section{Results}\label{sec:results}

\subsection{Potential energy surfaces}\label{sec:PES}
Table \ref{tab:PES} gives the coordinates and values of the minima of the PES's, and compares those to the existing literature values. Our approach is similar to earlier ones, with differences of the order of a few percent. Note that the geometry used for \ce{H2} might differ from one reference to another.  Importantly, the computed observables derived from both the intermediate precision and more refined PESs yield differences that remain below the 10\% targeted precision threshold for JWST. Indeed, we expect all the neglected effects, in particular non-rigid molecules and non-Born-Oppenheimer perturbations, to be of that order of magnitude, as discussed in the so-called ``platinum" approximation \citep{van_der_avoird_intermolecular_1994, 2019JChPh.151g0901K}.

\begin{table}[h!]
\centering 
\caption{Global minima of the $V_{\ce{H_2-CO2}}$ and $V_{\ce{He-CO2}}$ potentials. Comparison with the literature. Angles are identical for this work and the literature references. Distances in Bohr, angles in degrees, potential values in cm$^{-1}$.}\label{tab:PES}
\begin{tabular}{c|ccc|cc|cc}
\hline
System & \multicolumn{3}{|c|}{Angles} & \multicolumn{2}{c|}{This work} & \multicolumn{2}{c}{Literature} \\
& $\theta_1$ & $\theta_2$ & $\phi$& $r$ & $E$ & $r$ & $E$ \\
\hline
\ce{H2-CO2}& 90 & 90&  0& 5.595 &-222.65 & 5.612 & -219.75 (a)  \\
 & 90 & 90&  0& & & 5.6017 & -225.77 (b)  \\
 \hline
\ce{He-CO2}&   90 & -- &  -- & 5.77 & -48.55 & 5.78 &-49.22 (c) \\
\hline
\multicolumn{8}{l}{(a)  \cite{Li:2010aa} ; (b) \cite{hellman2025};}\\
\multicolumn{8}{l}{(c) \cite{2022JChPh.156j4303G} }\\
\hline
\end{tabular}
\end{table}

Fits (see section \ref{sec:theofit}), see \cite{Rist:2012aa}, were performed taking into account the symmetries of both H$_2$ and \ce{CO2}. For He collisions, $P_{\ell_1}(\cos\theta)$, $\ell_1=0,2,\ldots, 12$ were used. For H$_2$ collisions, we have $\ell_1=0,2,\ldots, 24$, $\ell_2=0,2, 4 ,6$ and total $\ell\leq 26$, resulting in 158 terms (see Eq. \ref{eq:fit1}). The quality of the fit for the $V_{\ce{H_2-CO2}}$ PES is discussed in detail in our previous work \cite{Wiesenfeld:2025aa}. The actual values are given in the data files (Appendix \ref{sec:tables}), in a format readable by the \YUMI\,code.

In the radial coordinate $r$, fits are interpolated for $r_{\text{min}}\leq r\leq r_{\text{max}}$ using cubic spline coefficients. For $r<r_{\text{min}}$, the potential is extrapolated by an exponentially increasing function. For large distance, inverse power functions $C_n/r^n, \, n\geq 6$ were used for the main terms, with a limit $|V|>0.01$ cm$^{-1}$.

The \ce{He-CO2} PES fit is precise, with an r.m.s. error of $\sim$0.01\% between $V_{\text{abinit}}$\, (see \autoref{fig:potential}) and $V_{\text{fit}}$. For the \ce{H2-CO2} PES fit, an r.m.s. error of 1\% is found. Lowering this error requires doubling the number of \abinit\, points, rendering the computation heavy.

\subsection{Dynamical computation results}

As described earlier, \autoref{sec:spy}, we need to compute the inelastic and elastic complex matrix elements of $\mathsf S(E)$ or $\mathsf T(E)= \mathsf 1 - \mathsf S(E)$ matrices. Previous work on ro-vibrational transitions has shown that, most of the time, inelastic rates are dominated by rotational inelastic, vibrationally elastic rates \cite{Wiesenfeld:2023aa}. While this is by no means a general conclusion, it is empirically supported here for \ce{CO2} molecules, as experiments have shown the pressure broadening depends very weakly on the vibration levels considered ($<2\%$) \citep{tan2022}.  We thus compute all matrix elements within the ground vibrational level, separating even and odd $j_1$ into two different series.

\begin{figure*}[ht!]
    \centering
    \includegraphics[width=1.0\linewidth]{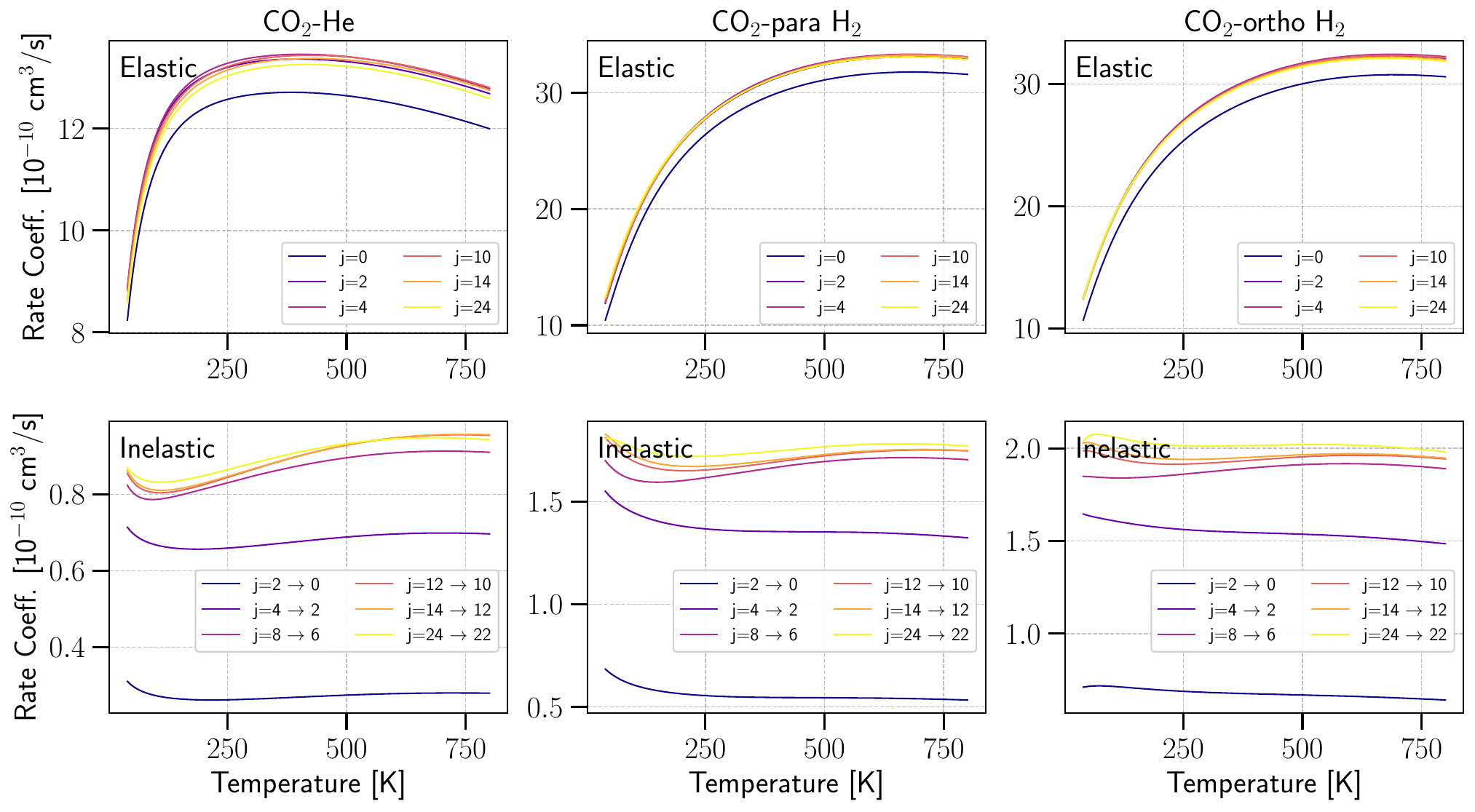}
    \caption{Elastic and inelastic rate coefficients $k(T)$ for ground-state CO$_2$ colliding with He and -H$_2$.  Elastic rate coefficients are roughly an order of magnitude larger than inelastic ones. \label{fig:rates}}
    
\end{figure*}

\subsubsection{Convergence criteria} 
\label{sec:convergence}
Convergence of the computations of cross-sections \( \sigma^{\text{PB;PS}}_{j_1''\leftarrow j_1'}(E)\) (Eq. \eqref{eq:section})  and associated rates is difficult to ensure, requiring trial and error for tuning computational parameters such as \textsc{nsteps} (see \autoref{sec:dynamics}) to carry out dynamical computation, given the scarce literature on the topic. Several parameters are to be set: \textsl{(i)} the total amount of \ce{CO2} rotational levels, which here amounts to the maximum value of $j_1$; \textsl{(ii)} the values of $j_2$, describing the rotational dynamics of \ce{H2}; \textsl{(iii) }the actual integration parameters for the Riccatti scheme (see \autoref{tab:param_dyn}), in particular $R_{\rm max}$; \textsl{(iv)} the largest value of total angular momentum $J$ in the sums of equations \eqref{eq:section}.  Also, \autoref{fig:convergenceFigure} shows the $J$ behavior of the convergence for the sums in Eq.\eqref{eq:section}. In order to converge the rates $k_{f\leftarrow i}(T)$, for $T\lesssim 800$~K, $E\approx 2300$~\wn\,must be reached. We computed the dynamics from $10\leq E \leq 2300 \mwn$, with steps increasing from $5 \;\text{to}\; 100\, \mwn$ .

\begin{table}[h]
    \centering
     \caption{Extreme values of the convergence parameters. Actual parameters for other cases were given intermediate values. Energies in \wn. }\label{tab:param_dyn}
    \begin{tabular}{l|c|c}
    \hline
      Parameter   & $E=100$; $|m|=2$ & $E=1900$ ; $|m|=24$ \\
      \hline
        $j_1^{\rm max}$ &   59 &  75 \\
        $R_{\rm max}$ &  50   &  50  \\
        $J_{\rm max}$ & 70   &  120 \\
        \hline        
    \end{tabular}
   
    \label{tab:parameters}
\end{table}

Properly computing dynamics with H$_2$ is substantially difficult. For ortho-H$_2$, limiting to  $j_2=1$ is a reasonable approximation, as we \emph{do not} explicitly include ro-vibrational transitions \citep{2012PhRvA..86b2705D}. In contrast, for para-H$_2$, an inclusion of the $j_2=2$ levels should be investigated. However, the number of coupled states for each $\left|j_1, j_2, j_{12}\right>$ level (Appendix, section \ref{sec:notations}) goes from 1 to 5, resulting in more than an order-of-magnitude increase in computational cost while approaching practical memory limitations. We therefore adopted an approximation frequently employed in the calculation for rotational inelastic rate coefficients \citep{faure_rotational_2024,wiesenfeld2022quenching}.

\begin{figure*}[t!]
   \centering
  \includegraphics[width=1.0\linewidth]{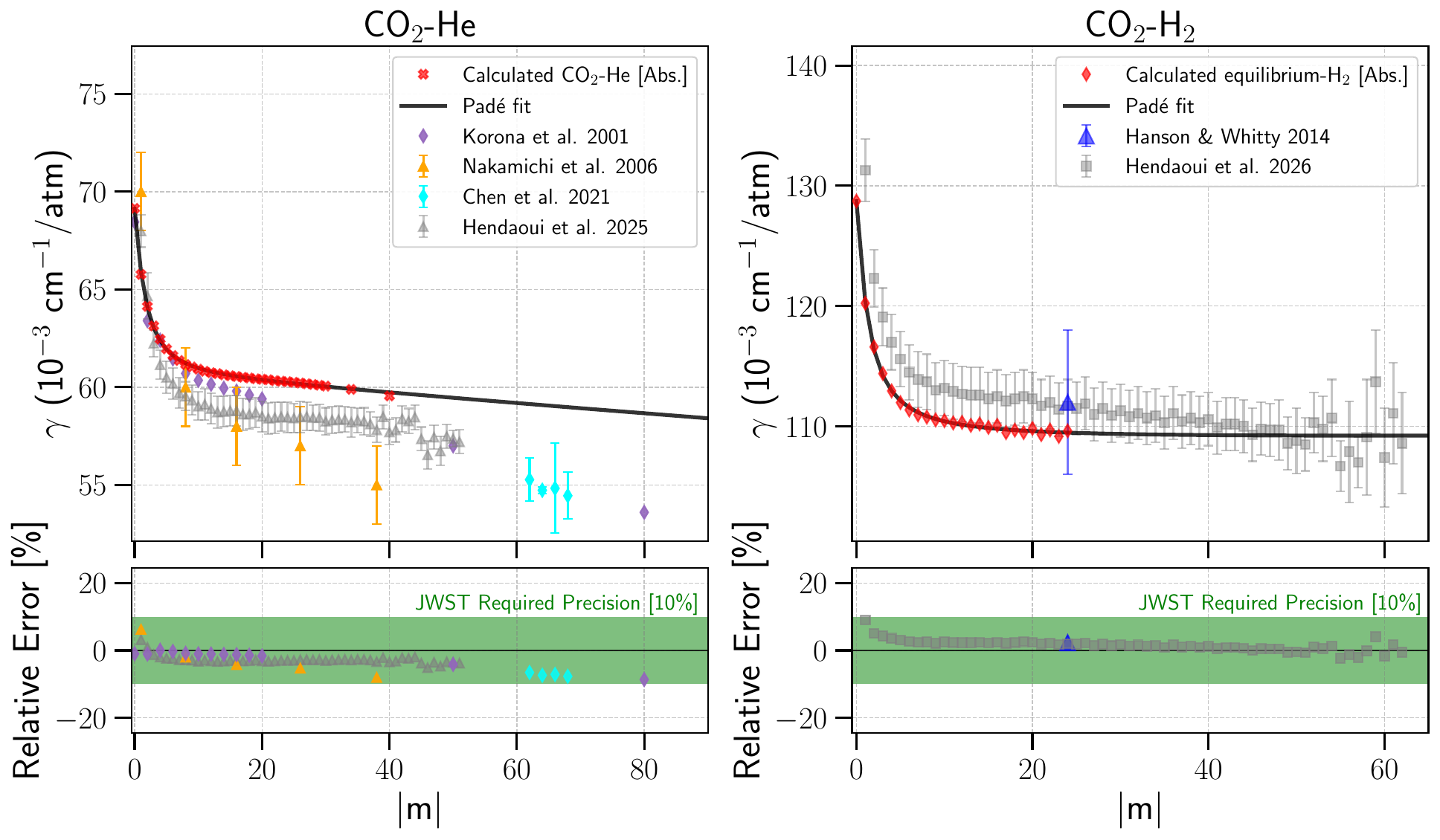}
     \caption{\textbf{Left:} Broadening \COTwo-He at 296 K as a function of rotational quantum number, compared with experimental measurements. The experimental values are taken from \citet{Hendaoui2025} and \citet{chen2021}, while the Pad\'e approximations are fitted to our fully ab initio calculations.  \textbf{Right:} Broadening \COTwo-\HTwo (equilibrium ortho/para mixture) at 296 K as a function of rotational quantum number, comparison with experiment. The experimental values are taken from \citet{hanson2014} and \citet{hendaoui:2026}. }
    \label{fig:PBvsM}
\end{figure*}

For the temperature range examined here ($T\gtrsim 40$ K), the detailed description of the $j_2>1$ scattering is no longer needed. We thus approximate all $j_2>0$ sections to the one with $j_2=1$, and use the relevant weight for the ratio $\rho(T)=n_{\ce{H2}(j_2>0)}(T) / n_{\ce{H2}(j_2=0)}(T)$. For {T=296 }K, at equilibrium, $\rho=0.866$. We use this ratio to present the results in \autoref{fig:PBvsM}. For the temperature dependence, a proper varying  $\rho(T)$ is employed. A full analysis of the $j_2$ dependence at moderate $j_1$ and $T$, for the $\ce{CO-H2}$ system, was used in \cite{Thibault:2000aa}, to calibrate pressure cross sections and pressure broadening coefficients. 

\subsubsection{Elastic and Inelastic scattering rates}
We computed, using \autoref{eq:secord}, the usual population transfer sections and rates, keeping in mind that only $j_1 \text{ odd/even} \leftrightarrow j_1 \text{ odd/even}$ are computed. Examples of rates (elastic and inelastic) are shown in \autoref{fig:rates}, for He, para-($j_2=0$) and ortho- H$_2$ ($j_2=1$) collisions. 
As is common, all rates connecting the $j_1=0$ level are particular, due to the degeneracy of the triangle conditions in the various recoupling 3-j or 6-j symbols appearing in the evaluation of the matrix element in \autoref{eq:secord}. 

The full sets of elastic and inelastic rates (in $10^{-10} \mathrm{\,cm^3\,s^{-1}}$) are provided as Supplemental Data in a \href{https://doi.org/10.5281/zenodo.20435057}{Zenodo repository}, alongside energy levels with respect to $j_1=0$ energy level taken at the origin. The rates are provided according to the LAMDA convention \citep{refschoier}, used in the recent EMAA (https://dx.doi.org/10.17178/EMAA) base, except that the Einstein $A$ coefficients are not included. Recall that we include neither ro-vibrational transfer rates nor $j'_1$ odd~$\leftrightarrow$~odd $j''_1$ transitions. These data could be used in helping to model the spectral line intensities in the IR region, but with utmost caution, since the rates of the $v''\leftarrow v'$ transitions and the timescales of the IR transitions are not taken into account, all the more that the purely $v'=v''$ transitions are forbidden for electric dipolar transitions. Direct comparison with literature population-transfer rates (\autoref{eq:secord}) is therefore of limited value. However, a cursory comparison of the data of \autoref{fig:rates} with those for other rod-like molecules, like CO (small dipole, \cite{Yang:2006aa}), or heavy-heavy diatomic, like SiO \citep{Balanca:2018aa}, shows good plausibility of our present results.

\subsubsection{Pressure broadening data}
\paragraph{Results}
\emph{Without need for any empirical scaling}, the present calculations reproduce the experimental measurements on an absolute scale, both in terms of temperature (or collision energy) and in the magnitude of the broadening coefficients.

\begin{figure*}[!htbp]
    \centering
    \includegraphics[width=1.02\linewidth,trim = 0cm 0.25cm 0cm 0.0cm, clip=true]{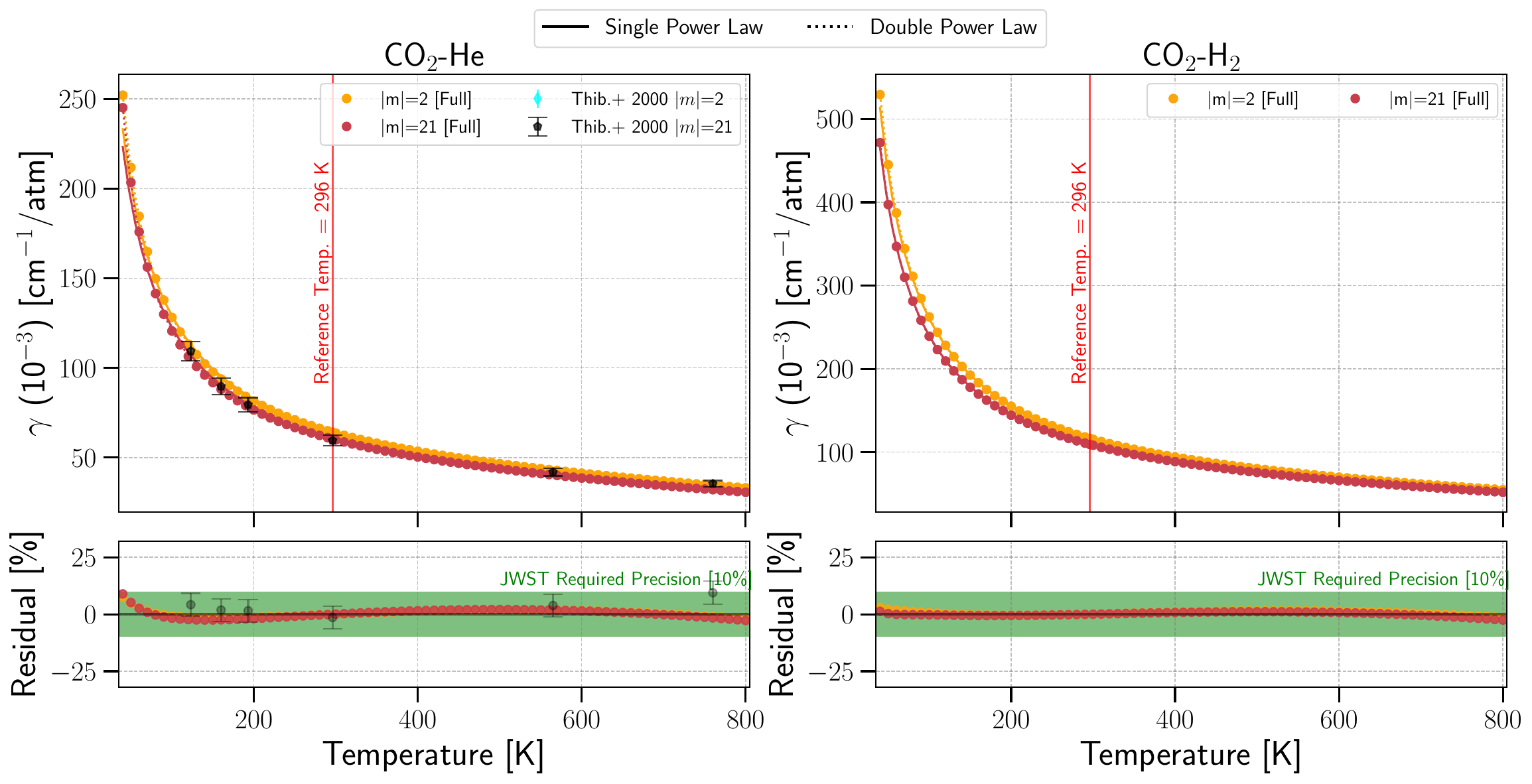}
    \caption{ Broadening of \COTwo-He~as a function of temperature for two transitions ($|m|$=11, 21) from two different methods for calculations and their corresponding fit in Single Power Law and Double Power Law, which shows good agreement with the values reported in \citet{Thibault:2000aa}. At both low and high temperatures, the single power-law (SPL) fit, shown as a solid line, exhibits larger deviations from the calculated values. Nevertheless, the discrepancies remain below the 10\% threshold typically required for JWST applications. The double power-law (DPL) fit, shown as a dotted line, provides a more accurate representation. Residuals, which show the relative difference between the fit values vs the calculated values, are only shown for the SPL for clarity.\label{fig:temperatureDependence}}
\end{figure*}

\paragraph{Fits}
We fit the temperature dependence of all pressure broadening coefficients over the range 40--800 K. Initially,  we fit a linear power law (also known as Single Power Law, henceforth SPL), parametrized as:
\begin{align}
    \label{eq:powerLaw}
    \gamma_{|m|}(T) &= \gamma_{|m|}(T_0)\cdot\left(\frac{T_0}{T}\right)^n 
\end{align}
where $\gamma_{|m|}(T_0)$ denotes the pressure broadening at the standard reference temperature of 296 K. However, to capture the behavior of pressure broadening over a wide range of temperature to the requirement of the precision of 10\% \citep{Wiesenfeld:2025aa}, a Double Power Law (henceforth, DPL) is often prescribed, as was discussed in the \citet{stolarczyk2020} and as can be seen in \autoref{fig:temperatureDependence}. We use the same formulation as Eq. 12 from \citet{stolarczyk2020}: 

\

\begin{align}
    \label{eq:doublepowerLaw}
   \gamma_{2, |m|}(T) &= g_2\cdot\left(\frac{T_0}{T}\right)^k + g_2'\cdot\left(\frac{T_0}{T}\right)^{k'},\ 
\end{align}
which has four independent parameters ($g_2$, $k$,  $g_2'$ \& $k'$) which are listed in \autoref{tab:gamma}. In order to break the degeneracy that exists between $g_2$, $k$,  $g_2'$, $k'$, while fitting the parameters, we enforce that $g_2$ is within 30\% of the reference broadening parameter ($\gamma(T_0)$) obtained for the SPL. The DPL offers a better fit to our calculated pressure broadening parameters compared to SPL, although the latter might be enough for JWST-related applications (see \autoref{fig:temperatureDependence}). For both  He and \ce{H2}, the power coefficients vary smoothly between the two computational approaches.

\begin{figure*}[h!]
    \centering
    \includegraphics[width=1.0\linewidth, trim = 0.0cm 0.25cm 0cm 0.0cm, clip=true]{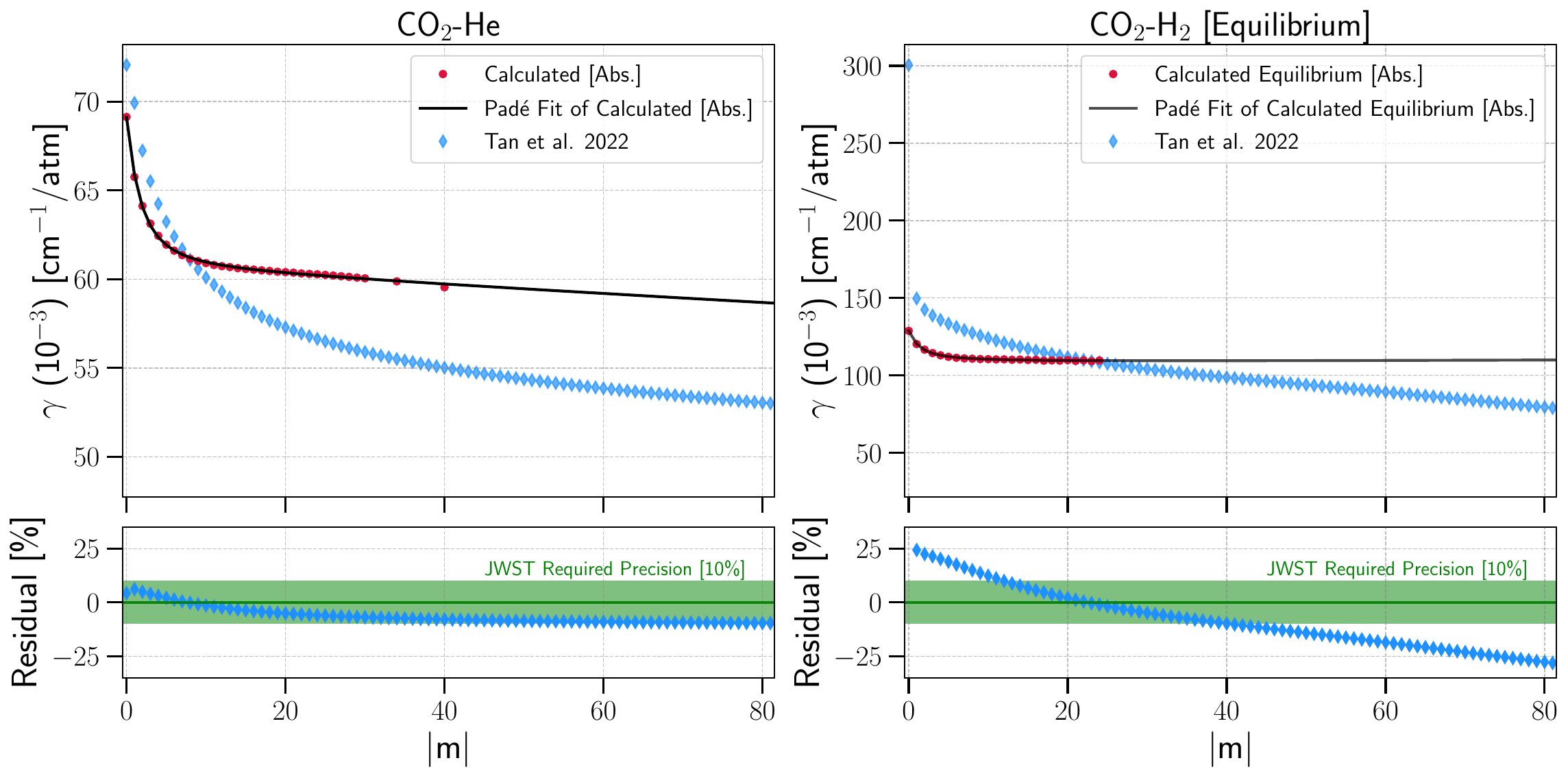}
    \caption{Fitting of Pad\'e approximation to our computed pressure broadening values at the reference temperature (296 K). We fitted a second-order Pad\'e approximation for both the broadening of helium and hydrogen, which are tabulated in \autoref{table:PadeCoefficientsTable}. The residuals shown in both bottom panels are the relative difference between the Pad\'e fit we obtained from our computation \textit{vs.} those suggested in \citet{tan2022} (and used in HITRAN2020 and 2024).\label{fig:padeApproximation}}
\end{figure*}

The $|m|$ dependence of is modeled with a Pad\'e approximation, defined as :
\begin{equation}\label{eq:pade}
    \gamma_{x}(\vert m \vert) = \frac{\sum\limits_{i=0}^k a_i |m|^{i} }{1 + \sum\limits_{j=1}^{k+1} b_j |m|^{j}}
\end{equation}
where, $a_0$ takes the value of $\gamma_{m=0}$. It is common to use third order (i.e. $k$=3) for fitting this relation \citep[e.g.][]{tan2022}. For our purposes, we adopt a second-order Pad\'e approximation. We fit the above expression to the values in the \autoref{table:PadeCoefficientsTable} and the fits are shown in \autoref{fig:padeApproximation}. These fits enable extrapolation to values beyond the calculated range while smoothing out any irregular features. However, such extrapolations should be treated with caution as they can produce unphysical values (at large $|m|$), as observations remain limited in this regime \citep{ngo2025}.

The full table of pressure broadening coefficients is given in the Supplementary data; see a description in \autoref{sec:tables}, and is available via \href{https://doi.org/10.5281/zenodo.20435057}{Zenodo\footnote{ https://doi.org/10.5281/zenodo.20435057}}.

\begin{deluxetable*}{ccccccc|ccccccc}[!htbp]
\vspace{-0.5cm}
\tablecaption{\label{tab:gamma}
Pressure broadening coefficients $\gamma$ at 296\,K for \ce{CO2}–He collisions and \ce{CO2}–\ce{H2} [Equilibrium]. $\gamma$ is in $10^{-3}\,\mwn/\mathrm{atm}$; $m$ defined as in \autoref{sec:notations}; $n$ as in Eq. \eqref{eq:powerLaw}; $k,\,k'$, $g_2,\,g'_2$ from Eq. \eqref{eq:doublepowerLaw}.}
\tablehead{
\multicolumn{7}{c}{\textbf{\ce{CO2}-He collision}} & \multicolumn{7}{c}{\textbf{\ce{CO2}-H$_2$ collision}} \\
\colhead{$|m|$} & \colhead{$\gamma$($T_0$)} & \colhead{$n$} & \colhead{$g_2$}& \colhead{$k$} & \colhead{$g'_2$} & \colhead{$k'$} &  \colhead{$|m|$} & \colhead{$\gamma$($T_0$)} & \colhead{$n$} & \colhead{$g_2$} &\colhead{$k$} & \colhead{$g'_2$} & \colhead{$k'$} 
}
\startdata
0 & 69.1340&0.6678&67.9539&0.6294&0.9602&1.9675 & 0 & 128.7336&0.7667&124.1466&0.7344&4.6866&1.4292\\
1 & 65.7560&0.6520&64.9492&0.6206&0.5591&2.1081 & 1 & 120.2356&0.7502&113.0030&0.7083&7.2925&1.2836\\
2 & 64.1240&0.6449&63.4019&0.6149&0.4637&2.1676 & 2 & 116.6056&0.7406&106.7702&0.6970&9.9163&1.1489\\
3 & 63.1260&0.6422&62.3759&0.6106&0.4852&2.1642 & 3 & 114.3931&0.7355&107.2273&0.6981&7.2382&1.2030\\
4 & 62.4440&0.6402&61.6586&0.6077&0.5177&2.1346 & 4 & 112.9479&0.7325&103.6381&0.6917&9.3914&1.1241\\
5 & 61.9450&0.6384&61.1535&0.6058&0.5328&2.1101 & 5 & 111.9773&0.7305&102.2863&0.6933&9.8288&1.0766\\
6 & 61.6080&0.6391&60.7543&0.6046&0.6006&2.0760 & 6 & 111.3681&0.7308&101.5314&0.6941&9.9905&1.0681\\
7 & 61.3570&0.6395&60.4889&0.6044&0.6297&2.0584 & 7 & 110.9609&0.7303&101.1789&0.6998&9.9982&1.0168\\
8 & 61.1510&0.6390&60.3126&0.6048&0.6038&2.0655 & 8 & 110.7721&0.7332&101.0257&0.6970&9.9343&1.0686\\
9 & 61.0220&0.6404&60.1795&0.6052&0.6158&2.0750 & 9 & 110.5379&0.7322&100.8548&0.6993&9.8622&1.0396\\
10& 60.9080&0.6404&60.0875&0.6057&0.5953&2.0862 & 10 & 110.4754&0.7329&100.8976&0.7051&9.8169&0.9997\\
11& 60.8000&0.6396&60.0536&0.6066&0.5203&2.1315 & 11 & 110.2882&0.7301&109.6665&0.7346&0.8746&0.1503\\
12& 60.7330&0.6401&59.9935&0.6067&0.5107&2.1519 & 12 & 110.3330&0.7328&100.6573&0.7108&9.9296&0.9471\\
13& 60.6830&0.6409&59.9273&0.6068&0.5207&2.1542 & 13 & 110.0864&0.7281&104.2603&0.7273&6.0680&0.7280\\
14& 60.6210&0.6407&59.9050&0.6072&0.4811&2.1923 & 14 & 110.2017&0.7327&100.4236&0.7064&9.9610&0.9762\\
15& 60.5740&0.6409&59.8507&0.6067&0.4795&2.2066 & 15 & 109.9159&0.7264&109.5039&0.7243&0.6002&0.7205\\
16& 60.5370&0.6412&59.8149&0.6066&0.4820&2.2126 & 16 & 110.0878&0.7322&101.8467&0.7070&8.3773&1.0067\\
17& 60.4980&0.6418&59.7496&0.6061&0.4988&2.2096 & 17 & 109.4841&0.7186&109.3634&0.7122&0.0010&4.2412\\
18& 60.4620&0.6424&59.7039&0.6059&0.5057&2.2187 & 18 & 109.6861&0.7250&109.6806&0.7201&0.0010&4.6773\\
19& 60.4180&0.6424&59.6382&0.6052&0.5197&2.2122 & 19 & 109.4848&0.7228&109.5005&0.7199&0.0021&3.3889\\
20& 60.3510&0.6523&59.6539&0.6197&0.2902&2.4403 & 20 & 109.8272&0.7316&109.1987&0.7200&0.6908&1.7624\\
21& 60.3180&0.6538&59.5922&0.6203&0.3027&2.4333 & 21 & 109.3159&0.7239&109.2852&0.7188&0.0143&3.3107\\
22& 60.2660&0.6542&59.5471&0.6209&0.2891&2.4518 & 22 & 109.6973&0.7326&109.1980&0.7190&0.5259&2.0309\\
23& 60.2300&0.6552&59.4849&0.6213&0.2992&2.4395 & 23 & 109.1654&0.7251&109.0525&0.7164&0.0652&2.9162\\
24& 60.1950&0.6566&59.4382&0.6223&0.3047&2.4372 & 24 & 109.6284&0.7344&108.9786&0.7174&0.6308&2.0611\\
25& 60.1420&0.6571&59.3797&0.6230&0.2975&2.4441&\nodata&\nodata&\nodata&\nodata&\nodata&\nodata&\nodata\\
26& 60.0930&0.6577&59.3215&0.6237&0.2943&2.4433&\nodata&\nodata&\nodata&\nodata&\nodata&\nodata&\nodata\\
27& 60.0440&0.6587&59.2639&0.6245&0.2957&2.4408&\nodata&\nodata&\nodata&\nodata&\nodata&\nodata&\nodata\\
28& 59.9860&0.6594&59.1950&0.6252&0.2941&2.4375&\nodata&\nodata&\nodata&\nodata&\nodata&\nodata&\nodata\\
29& 59.9200&0.6596&59.1302&0.6261&0.2811&2.4470&\nodata&\nodata&\nodata&\nodata&\nodata&\nodata&\nodata\\
30& 59.8830&0.6592&59.0950&0.6260&0.2790&2.4461&\nodata&\nodata&\nodata&\nodata&\nodata&\nodata&\nodata\\
34& 59.8770&0.6402&58.8511&0.5984&0.7670&2.0393&\nodata&\nodata&\nodata&\nodata&\nodata&\nodata&\nodata\\
40& 59.5430&0.6330&58.7036&0.5978&0.5894&2.0735&\nodata&\nodata&\nodata&\nodata&\nodata&\nodata&\nodata
\enddata
\end{deluxetable*}

\begin{deluxetable*}{rcc|cc}[!ht]
\tablecaption{\label{table:PadeCoefficientsTable}Pad\'e Approximations shown in \autoref{fig:padeApproximation}.}
\tablehead{
 & \multicolumn{2}{c}{\ce{CO2}-\ce{He}}  & \multicolumn{2}{c}{\ce{CO2}-H$_2$} \\
\colhead{Parameter} &
\colhead{\citet{tan2022}}&
\colhead{This Work} & 
\colhead{\citet{tan2022}}&
\colhead{This Work}
}
\startdata
\hline
$a_0$ & 7.2060 $\times 10^{-2}$ & 6.9114 $\times 10^{-2}$ & 3.0051 $\times 10^{-1}$ & 1.2877 $\times 10^{-1}$ \\
$a_1$ & -2.2690 $\times 10^{-2}$ & 3.1884 $\times 10^{-2}$ & 1.9992  & 1.3510 $\times 10^{-1}$ \\
$a_2$ & 1.0172 $\times 10^{-1}$ & 7.8119 $\times 10^{-3}$ & -2.8360 $\times 10^{-2}$ & 6.6474 $\times 10^{-2}$ \\
$a_3$ & 1.1680 $\times 10^{-2}$ & \nodata & 6.3494 $\times 10^{-4}$ & \nodata \\
$b_1$ & -3.2460 $\times 10^{-1}$ & 5.2285 $\times 10^{-1}$ & 1.4150 $\times 10^{1}$ & 1.1293  \\
$b_2$ & 1.4333  & 1.2892 $\times 10^{-1}$ & 2.7310 $\times 10^{-2}$ & 6.1383 $\times 10^{-1}$  \\
$b_3$ & 2.1907 $\times 10^{-1}$ & 5.5926 $\times 10^{-5}$ & -9.2860 $\times 10^{-5}$ & -9.8419 $\times 10^{-5}$  \\
$b_4$ & 8.9402 $\times 10^{-5}$ & \nodata & 6.2540 $\times 10^{-5}$ & \nodata  \\
\enddata
\end{deluxetable*}

\begin{figure*}[ht!]
   \includegraphics[width=1.025\linewidth, trim = 0.0cm 0.25cm 0cm 0.0cm, clip=true]{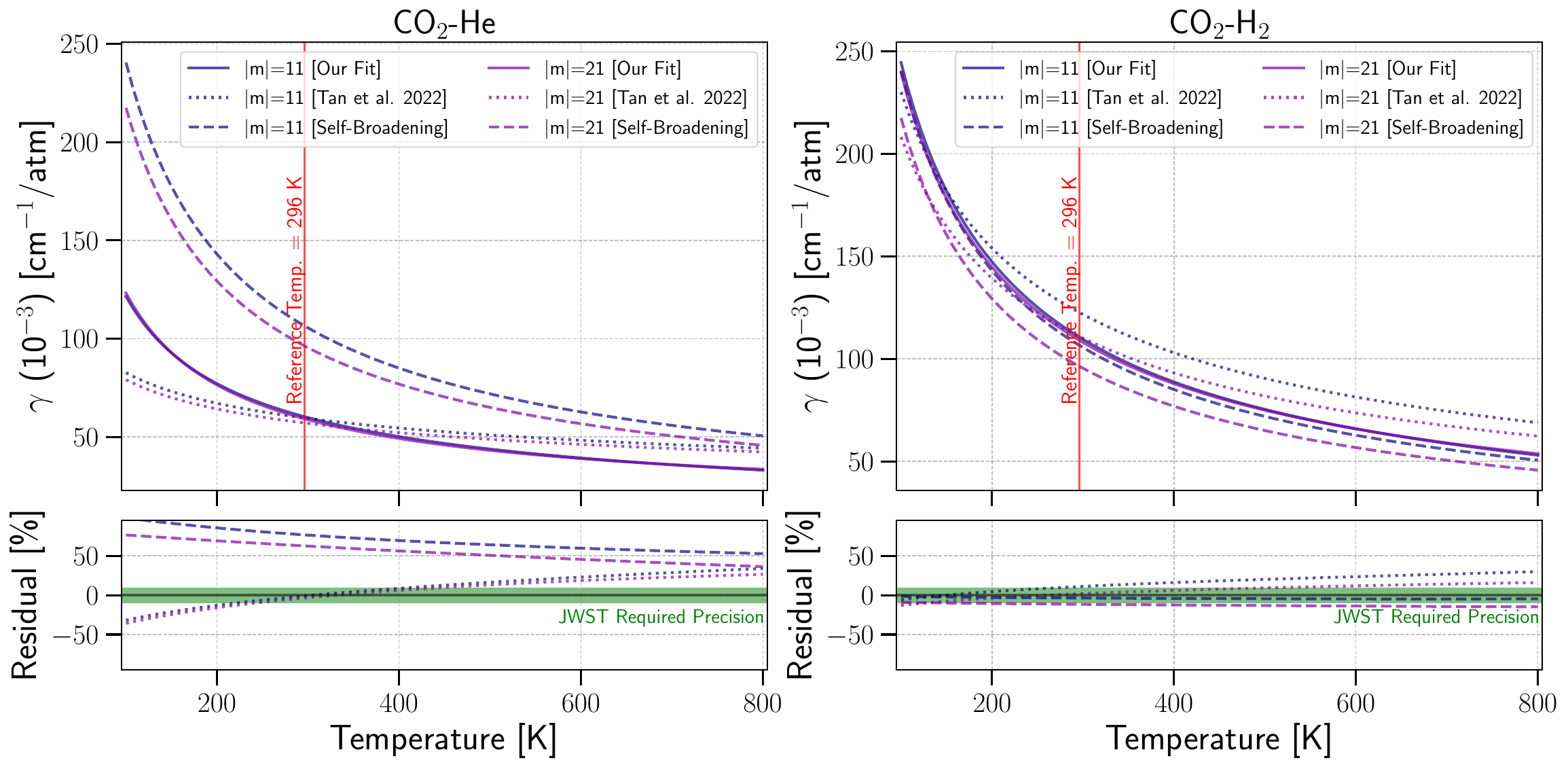}
    \caption{Comparison of the pressure broadening coefficients for two transitions ($|m|$=11, 21) across 40 K to 800 K in comparison to previous state-of-the-art estimations [HITRAN]. By refining the relevant power coefficients, we achieve the required improvements in the accuracy of pressure broadening coefficients to power instrument-limited exoplanetary studies in the \textit{JWST} era. This contrasts with the parameters available before the present calculations (for \ce{CO2}-\ce{H2}), where discrepancies reached up to 2$-$5$\times$ the precision requirement (typically for $T\gtrsim 400$\ K). Comparisons with self-broadening are presented to highlight that their use in place of broadening by hydrogen (an approximation used in previous exoplanetary studies) is not an adequate substitute in the era of \textit{JWST}.  \label{fig:comparison}}
\end{figure*}

\section{Comparison to experimental Values and HITRAN.}
\label{sec:compare}
Our calculations reproduce the observed dependence on $|m|$ (see \autoref{fig:PBvsM}) for both collision systems without the need for empirical scaling. For \ce{CO2-H2}: the \citet{hanson2014} measurement targeted P(24) transition of the 20012-00001 band at different temperatures, whereas \citet{padmanabhan2014} covers P(16)-P(34) transitions of the 30012-00001 band at room temperature (see \citet{hitran2020} for spectroscopic notations).  Since very few transitions were available for \ce{CO2-H2} broadening, recent editions of the HITRAN database \citep{hitran2020, HITRAN2024} have adopted scaled air-broadening parameters \citep[as described in][]{tan2022} as they are very comprehensive thanks to numerous studies \citep[e.g.,][]{10.1016/j.jqsrt.2020.107283}, with a scaling factor derived from P(24) transition values reported by \citet{hanson2014}. The reported temperature dependence coefficient for the same transition of 0.58 was adopted universally for all transitions of \ce{CO2} in HITRAN. As shown in \autoref{fig:padeApproximation},  the rotational dependence of broadening due to air differs from that due to hydrogen, although in general, the aforementioned scaling used in HITRAN was justified considering the lack of data at the time \citep{tan2022}.  Our calculation now provides a more comprehensive estimation of the temperature dependence coefficient over a wide range of temperature and $|m|$. 

A new study of the \ce{CO2-H2} collision has been published very recently during the revision process of this paper, \cite{hendaoui:2026}, reporting both experimental and theoretical calculations. The experimental data are extensive and are shown in the relevant \autoref{fig:PBvsM}. We compare extremely well with the published data for most of the values of $|m|$ and $T$ (see \autoref{fig:PBvsM}). The agreement extends to the inferred temperature exponent of pressure broadening. While measurements were performed at room temperature, they computed the temperature dependence and fit the results. For SNL,  their typical coefficient is $n\sim 0.70 \pm 0.13$, close to our value of $n \sim 0.73$, a 4\% difference.

The calculations in \citep {hendaoui:2026} were performed with a very different approach, based on sampling of multiple classical trajectories. The potential energy surface used in that reference is also based on a very different approach, with high precision and much fewer \textsl{ab initio} points \citep[see][]{hellman2025}. Their published calculations and our fully quantum results achieve a similar level of accuracy relative to experiment, with all discrepancies remaining within the expected uncertainties. This agreement lends support to leveraging fully ab initio methodologies for determining absolute pressure broadening coefficients, assuming a limited dependence on vibrational state and transition \citep{hitran2020}.

On its side \ce{CO2-He} has been quite extensively studied. A detailed study of this system was performed by Thibault et al. group in their various works \citep{Thibault:2000aa, Korona2001}, showing an excellent match between the theoretical and experimental findings. As for the HITRAN database, the experimental broadening coefficients ($\gamma_{\mathrm{He}}$) for the 30013-00001 band of CO$_2$ in the 1.6 $\mu$m region were taken from~\citep{10.1039/b511772k}, who carried out measurements using continuous-wave cavity ring-down spectroscopy. These data were fit to the Pad\'e approximant, which was used to populate the entire database, ignoring the vibrational dependence (see \citet{tan2022}, for details). Following the addition of this model to HITRAN2020 for the $\gamma_\mathrm{He}$ half-widths of CO$_{2}$, new measurements became available~\citep{chen2021, Hendaoui2025} for the $\nu_3$ band of CO$_2$, covering a much larger range of rotational quanta. Due to the quality and comprehensiveness of \citet{Hendaoui2025}'s data, we use it to compare our calculations. In general, as can be seen from \autoref{fig:padeApproximation}, the agreement with the experimental data is very convincing, even if we tend to be systematically higher, by $\lesssim 5 \%$. At this level of precision, it is difficult to ascertain any definitive reason for this systematic error.

The temperature dependence for \ce{CO2}-\ce{He} was studied experimentally in  \cite{10.1039/b511772k, Deng2009, Brimacombe83, Thibault:2000aa}. When compared to the experimentally reported values in \citet{Thibault:2000aa}, which have measurements up to 760 K (comparable to our 800 K) our results largely match, as can be seen in \autoref{fig:temperatureDependence}. This indicates improvement over the current use of temperature exponents in the current HITRAN database, which adopted values using a piecewise linear law based on data from \citet{Brimacombe83, 10.1039/b511772k, Deng2009}. While our RPA estimates seem to agree more with the experimental values, it was previously observed that full calculation often provides more precise estimates at lower temperatures \citep{Faure:2013aa}.  As discussed in previous literature (such as \citet{Deng2009, Hendaoui2025}), relying on experiments alone is challenging because the experimental values do not always agree with one another for \ce{CO2-He}, primarily because the derived parameters depend on the adopted line-shape model. The recent calculations by \citet{HendaouiICARUS} extend to temperatures as high as 3000 K and $|m|$ close to 240. However, the absence of experimental values in this regime makes it difficult to validate the calculations. 

However, for comparison at standard temperature and pressure, given the close agreement with the recent experimental measurements (e.g. \citet{Hendaoui2025}), a $\pm 5 \%$ error bar on our values is rather a conservative estimate of the uncertainty. As shown in \autoref{fig:temperatureDependence}, \YUMI\, ultimately aims to yield the collisional parameters that achieve the targeted 10\% precision requirement for exoplanetary sciences with \textit{JWST}. Our calculations enable estimating the collisional properties for a wide range of $|m|$ and temperature spanning the parameter of interest for exoplanets, marking almost 10$\times$ improvement compared to what was previously used in exoplanetary retrieval models.

\section{Conclusion}\label{sec:conclusion}

We have computed pressure broadening $\gamma$ coefficients and elastic/inelastic rate coefficients for collisions of CO$_2$ with He and H$_2$. We characterize both the rotational dependence, expressed through the $|m|$ index, and the temperature dependence of these collisional properties over 40--800~K. The resulting pressure broadening coefficients reproduce available experimental measurements on an absolute scale, without empirical correction factors, and meet the $\sim$10\% precision requirement identified for \textit{JWST}-era exoplanet atmospheric studies. This demonstrates that fully \abinit\ quantum-scattering calculations can provide reliable, database-ready broadening parameters for systems where experimental data are sparse or difficult to obtain. 

We provide parametrization comparable to those used in HITRAN, including temperature-dependent fits and Pad\'e approximations as a function of rotational quantum number. These results represent a substantial improvement over previously available approximations for CO$_2$--H$_2$, especially at $T>400$~K, where existing parameters fall outside the desired precision by 2-5$\times$. Extending fully \abinit\ close-coupling calculations to very high temperatures and very high rotational quantum numbers ($j_1\lesssim250$), as required for some HITEMP applications, is not currently realistic at this level of theory. The framework presented here therefore provides an absolute-scale anchor for more approximate methods, surrogate models, and extrapolation schemes needed at $T\gg800$~K \citep{HITEMP-CO2}.

Our approach relies on \textsl{ab initio} quantum chemistry calculations performed with \textsc{Molpro}~\citep{MOLPRO} and on the \YUMI\, quantum-dynamical framework, whose computational flow is summarized in \autoref{fig:flowcomput}. \YUMI\, was developed to enable large-scale quantum-scattering calculations on modern HPC platforms, using parallel execution and memory-management strategies suitable for the large coupled-channel systems encountered here. Ongoing development also focuses on exploiting GPU-accelerated linear algebra to further improve scalability. Details of \YUMI\, will be published elsewhere. Further algorithmic and computational optimizations, including accelerated numerical kernels and surrogate-model approaches, are being pursued to extend this framework to more complex collisional systems.

These results are to be put in parallel to low-energy scattering resonances,  performed, e.g. by \cite{Bergeat:2020aa,bergeat2022near,kuijpers_imaging_2025}. Both series (detailed in energy and/or absolute in section values) are results of a full quantum mechanical approach, encompassing all symmetries and principles of quantum mechanics. In the scattering regimes --as opposed to the spectroscopy-- these are stringent tests of our \abinit\ approach, and, ultimately, of all the well-known quantum mechanical formalisms that were to be used.

We also show that, for CO$_2$, the vibrational dependence of the $\gamma$ coefficients is smaller than the precision of both experiment and theory. This follows from the weak dependence of the CO$_2$ geometry on vibrational state, in contrast to many H-containing molecules, especially H$_2$O. Together with the expected sub-percent isotopic dependence \citep{10.1016/j.jqsrt.2024.109271}, this suggests that the present calculations can potentially be used for all twelve stable isotopologues of carbon dioxide in HITRAN. These results can also support HITEMP \citep{HITEMP-CO2}, ExoMol \citep{ExoMol2024}, AI-3000K \citep{AI-3000K}, CDSD \citep{CDSD-2024-PI}, and other databases that provide rotational quantum information for CO$_2$ lines.

Scaling first-principles opacity generation to more complex target and projectile molecules requires furthermore, traceable, expert-validated workflows for deriving, implementing, and validating increasingly complex close coupling formalisms. In this work, AI-assisted tools were used as aids for comparing published formalisms, identifying notation and normalization inconsistencies, supporting code review and debugging, and automating workflow organization, including handling the tens of thousands of generated files. The computational framework, physical assumptions, numerical implementation, and scientific results reported here were developed, evaluated, and validated by the authors.

\software{\textsc{Molpro} \citep{MOLPRO}, \YUMI\, (Jaïdane et al. in prep); public versions of ChatGPT, Gemini, and Anthropic Claude, used as AI-assisted tools for code review, debugging, workflow automation, and cross-checking notation and normalization conventions.}

\textit{Zenodo repository:} The full table of pressure broadening coefficients is given in the Supplementary data at the following Zenodo repository ({\url{https://doi.org/10.5281/zenodo.20435057}}).

\section*{Acknowledgements}
The authors acknowledge helpful discussions with Frances Gomez.  PN, LW, JdW, IG, and RH acknowledge NASA XRP grant 80NSSC25K7168. NJ and PN acknowledge support from the Université-Paris-Saclay. JK, DW, and CL were supported in part by Department of the Air Force Artificial Intelligence Accelerator and was accomplished under Cooperative Agreement Number FA8750-19-2-1000. A major part of the dynamical computations was performed using the MIT SuperCloud (\cite{reuther2018interactive}). Ab initio computations (software MOLPRO) and some dynamical computations were performed on the Jean-Zay IDRIS-CNRS supercomputer under contract A0140810769.  The authors acknowledge IDRIS and the MIT SuperCloud and Lincoln Laboratory Supercomputing Center for providing resources (HPC, software expertise, database, consultation) that have contributed to the research results reported here.  The authors wish to acknowledge the contributions and support of: Anderson, Arcand, Bergeron, Birardi, Bond, Bonn, Byun, Burrill, Edelman, Fisher, Gadepally, Gottschalk, Hill, Houle, Hubbell, Jananathan, Jones, Leiserson, Luszczek, Malvey, Michaleas, Milechin, Milner, Mohindra, Morales, Mullen, Pentland, Perry, Pisharody, Prothmann, Prout, Rejto, Reuther, Rosa, Ruppel, Rus, Sherman, Yee, Zissman.
The views and conclusions contained in this document are those of the authors and should not be interpreted as representing the official policies, either expressed or implied, of the Department of the Air Force or the U.S. Government. The U.S. Government is authorized to reproduce and distribute reprints for Government purposes, notwithstanding any copyright notation herein.

\clearpage

\appendix

\restartappendixnumbering 
\section{Notations}\label{sec:notations}
The angular momenta involved in the collision process are labeled as follows: $j_1$, \ce{CO2} (target) angular momentum; $j_2$, \ce{H2} (projectile) angular momentum; $\ell$, the orbital angular momentum of the projectile with respect to the target in the laboratory frame. All combine to yield $J$, the total angular momentum of the collision in the laboratory frame, a conserved quantum number. The coupling scheme of the angular momenta is as follows (not writing the magnetic quantum numbers because of overall rotational invariance): $|j_1>,\,|j_2>$; $ |j_1 >\otimes |j_2> \mapsto |j_{12}>$; $|j_{12}> \otimes |\ell>\mapsto |J>$. We characterize the IR transition by the (pseudo-) quantum number $m$, which is $m=-k$, for the $P(k)$ transition and $m=k$ for the $R(k)$ transition. There is no $Q$ branch here.

The kinetic energy of the collision (sometimes called the collision energy) is denoted by $E$. The total energy is $\mathcal{E}=E+E_{1}+E_{2}$, where $E_1$ and $E_2$ are the internal energies (rotations, vibrations, \ldots) of the target $1$ and the projectile $2$.  Here, as discussed in \autoref{sec:spy}, we do not add the vibrational energies to the internal energies. $E_1$ and $E_2$ are rotational energies only. 
The wavenumber associated with the collision is $k = \sqrt{2\mu E}/\hbar$, $\mu$, reduced mass of the collision.

\section{Details of PES computation}
\label{sec:fits}
\subsection{Geometries}
\begin{figure}[h]
    \centering
  \includegraphics[width=0.65\linewidth]{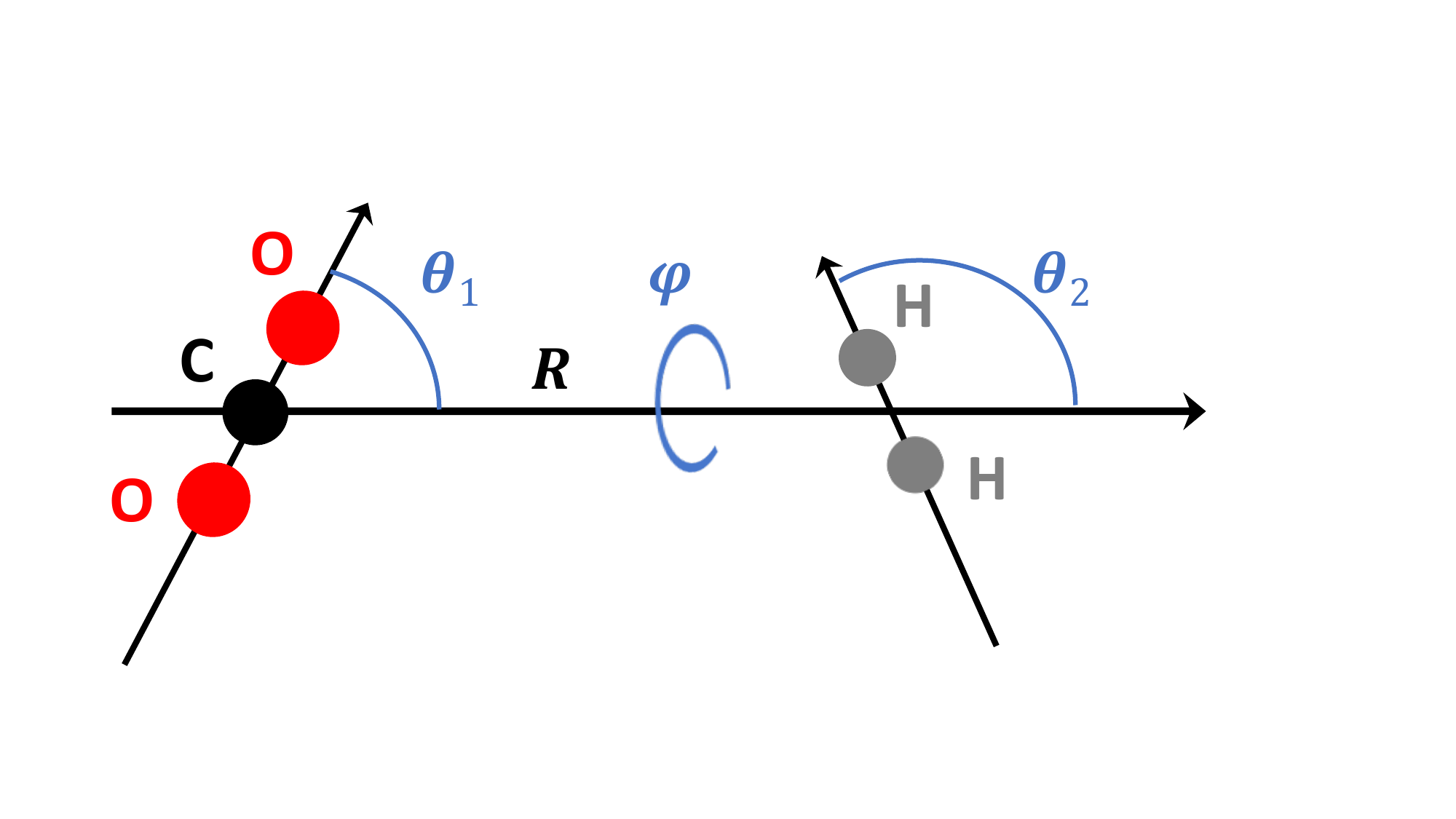}
    \caption{Geometry of the \COTwo--\HTwo\,complex.}
    \label{fig:geometry}
\end{figure}

The geometry of the \ce{CO2}-\ce{H2} colliding systems is presented in \autoref{fig:geometry}. It follows the conventions of \cite{GREEN:1975ae}, which is \emph{not} identical to the rotator-target collisions conventions used in \citet{valiron2008}.

\subsection{Extrapolation to complete basis set}
The basis sets used were the aug-CC-pV$X$Z type, from the MOLPRO basis set repository. We took $X=3(T),4(Q),5, \,\text{and }6$ basis sets \citep{pritchard_new_2019}. The extrapolation to the complete basis set was performed on the correlation energies, thanks to the  formula:

\begin{equation}\label{eq:cbs}
    E_{\mathrm{CBS}}= \frac{E_X^X -E_{X-1}^{X-1}}{X^X-(X-1)^{X-1}}
\end{equation}
with $X=4,5$. The Hartree-Fock contribution depends very weakly on the $X$ values, we used the highest $X$ available. We did use the aug-CC-pV6Z basis only to verify our extrapolation. All the results were in agreement to better than $0.5\,\mathrm{cm^{-1}}$ in the well region.

A recent study by \citet{hellman2025} performed an extremely precise computation of several van der Waals dimers, at more or less the platinum standard of \cite{2019JChPh.151g0901K}. It included complexes with H$_2$, O$_2$ and N$_2$ for significant molecules for the HITRAN database. They show, among other points, that high order corrections, including relativistic corrections, core correlations and extensions towards CCSDT(Q) changes the short range potential, in a maybe significant way for higher energy elastic scattering, up to a few percent, at $E\simeq 500 \mwn$. Note, however, that the geometries are not optimized, which might result in other non-negligible effects too, at this level of precision.

\subsection{Fitting functions and coefficients for potential}
We fit the PES's using the standard least-squares minimization (see \autoref{eq:fit1} and \cite{Wiesenfeld:2025aa}).

\begin{figure}[h!]
    \centering
 \includegraphics[width=0.75\linewidth]{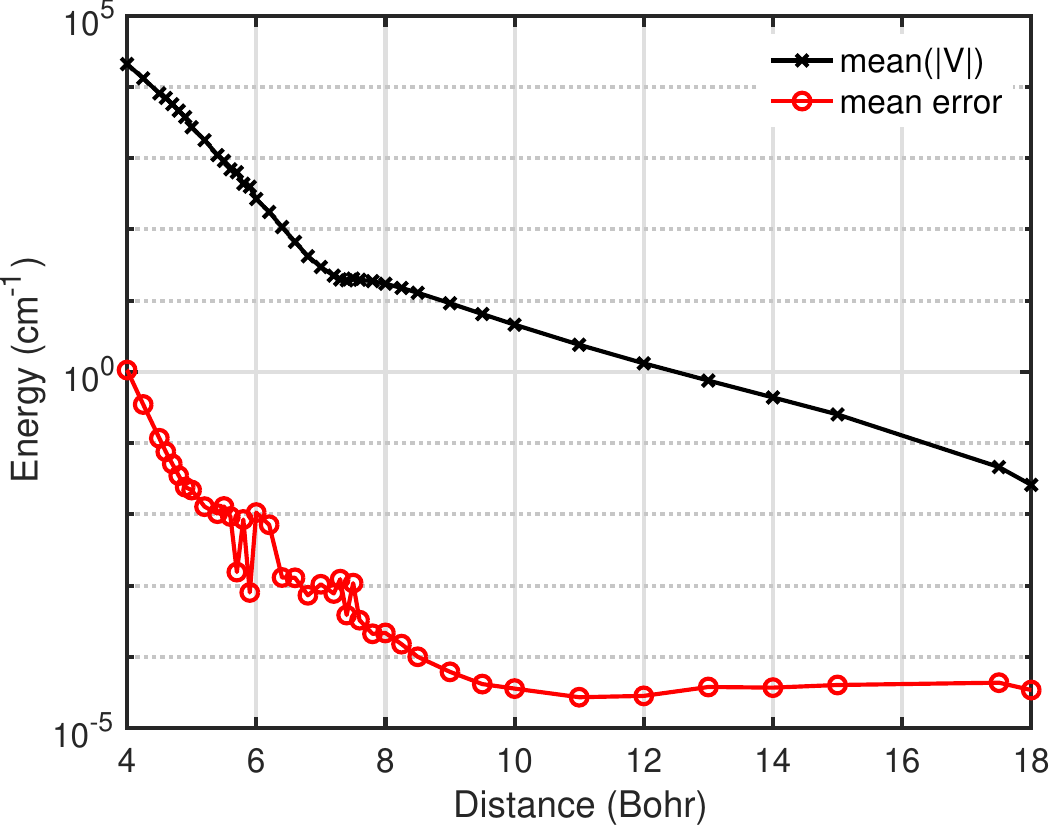}
    \caption{ Quality of the fit for the $V_{\ce{He-CO2}}$ PES. Shown is the average of $\left|V(r,\theta)\right|$ over the angle $\theta$ and the average over $\theta$ error , $\epsilon=|V_{\rm fit}- V_{\rm ab\,initio}|/|V_{\rm ab\,initio}|$.}
    \label{fig:potential}
\end{figure}

\section{Dynamics}
\label{sec:dynamics}
\subsection{Convergence of the  dynamics}
The integrator used in the dynamics \citep{manolopoulos_improved_1986} requires three main parameters, namely, the initial point (\textsf{RI}), the final point (\textsf{RF}), and the step (\textsf{STEP}) (the version used was not with variable step). We kept \textsf{RI} within the inner forbidden region (\textsf{RI} $<$4), \textsf{RF} well into the weak coupling region (\textsf{RF}$\geq$50) - distances in Bohr. \textsf{STEP} is given as a fraction of the de Broglie wavelength at a large distance. The standard settings are \textsf{STEP}=10 for high energies and \textsf{STEP}=100 for low energies  (here, $E\leq 10 \;\mathrm{cm^{-1}}$). The rotational basis set of \ce{CO2} is taken such as at least 3 levels are closed, with energies $|E-E_{j_1}| > V_{\rm min}$, where $|V_{\rm min}|$ is the minimum value of the potential, taken at 250  and 60 cm$^{-1}$ for \ce{H2} and He targets, respectively.

The para-H$_2$ modification was limited to $j_2=0$ and the ortho-H$_2$ to $j_2=1$. We tried some computations with $j_2=0,2$ but they proved intractable at $E\gtrsim 200\;\mathrm{cm^{-1}}$, $j_1\gtrsim 20$ because of the length of the computation and the size of the \textsf{T} matrix and the subsequent pressure broadening computation.

\begin{figure}[h]
    \centering
    \includegraphics[width=0.95\linewidth]{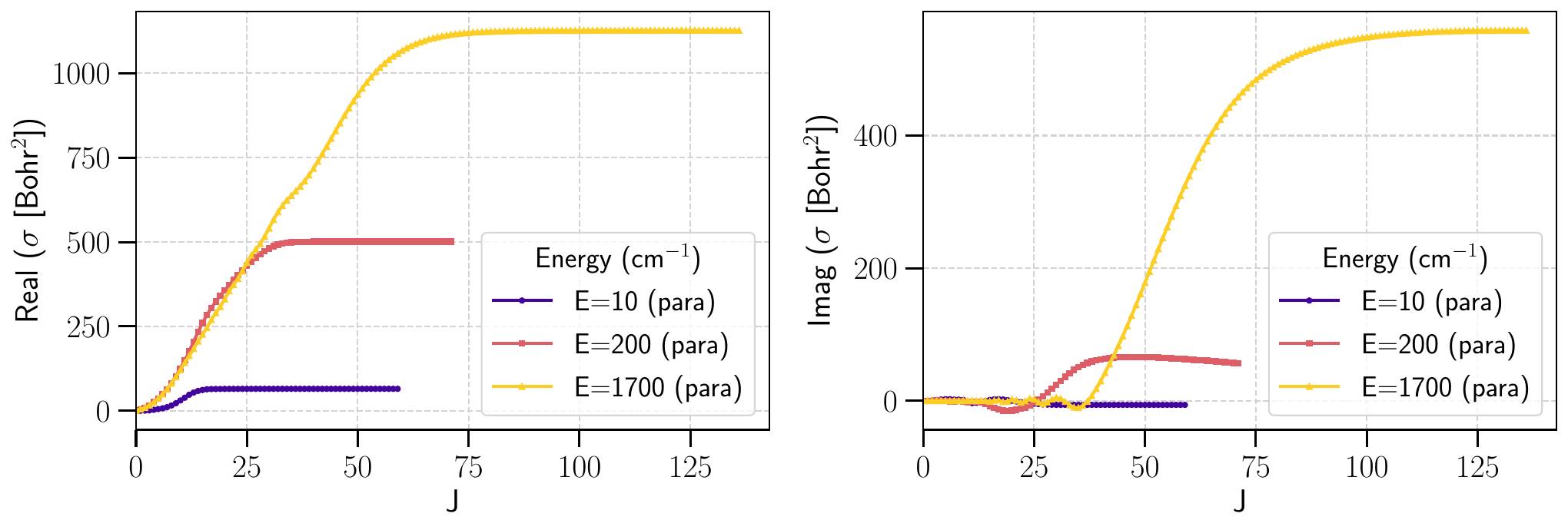}
    \caption{Convergence of the pressure broadening cross section for the \ce{CO2-para H2} corresponding to different kinetic energies for transition corresponding to $m=+10$ transition. The cumulative sum of the real and the imaginary part (which can be either positive or negative)  of the \autoref{eq:section} is plotted against the total angular momentum ($J$). }
    \label{fig:convergenceFigure}
\end{figure}

The next convergence parameter is the maximum value of the total angular $J$. While the convergence of the ordinary cross section \autoref{eq:secord} is easy to test and achieve, the convergence of the pressure broadening is not (and all the more for the pressure shift, which is additionally complicated by its vibrational dependence \citep{gamache2013} and absence of reliable experimental benchmarks). \autoref{fig:convergenceFigure} shows the convergence obtained for representative cases, with collisions with He and \ce{H2}:

\subsection{Convention for the rotator eigenfunctions:\\ Normalization of Legendre polynomials and Legendre associated functions}\label{sec:norm_legendre}
Depending on the authors, several normalizations of Legendre polynomials and associated functions have been used. Thus, we clarify the normalization scheme used in our framework.

Knowing that the spherical harmonics  obey the following normalization equation (stemming from the definition of rotor eigenfunctions) :
\begin{equation}\label{eq:ylm_norm}
\int_{\theta=0}^\pi \int_{\phi=0}^{2\pi} Y^*_{\ell m}(\theta, \phi) Y_{\ell' m'}(\theta, \phi)\,\sin\theta\mathrm d\theta \,\mathrm{d}\phi = \delta_{\ell \ell'}\delta_{m m'},
\end{equation}
The following normalizations and phases have been used. Note that existing scattering software and special function software may use different incompatible normalizations or phase conventions, but final observables should, evidently, be equal. We follow closely the \textsc{Molscat} conventions,  \citep[ch 1.3]{Zare:1988aa}, 

We have the following definition in
\begin{equation}
    Y_{\ell m}(\theta,\phi)=\mathcal{P}_\ell^m(\theta)\,\Phi_m(\phi)
\end{equation}
 The normalizations are 
 \begin{equation}
 \int_0^{2\pi}\Phi^*_m(\phi)\Phi_{m'}^{}(\phi)\mathrm{d}\phi=\delta_{mm'}
 \end{equation}
 and 
 \begin{equation} \label{eq:norm2}
 \int_0^{\pi}\mathcal{P}^*{}_{\ell}^m(\theta)\mathcal{P}_{\ell'}^{m}(\theta)\sin\theta\mathrm{d}\theta=\delta_{\ell \ell'}\quad .
 \end{equation}
Keeping in mind \autoref{eq:ylm_norm}, we have the following explicit forms:
\begin{equation}
    \Phi_m(\phi) = \frac{1}{\sqrt{2\pi}}\exp({im\phi})
\end{equation}
and:
\begin{equation}\label{eq:legendre_assoc}
\mathcal{P}_\ell^m(\theta)=(-1)^m\,\left[ \frac{2\ell+1}{2}\frac{(\ell-m)!}{(\ell+m)!}\right]^{1/2}P_\ell^m(\cos\theta)
\end{equation}
where $P_\ell^m(\cos\theta)$ are associated Legendre functions, whose normalization is given by \autoref{eq:norm2} and \autoref{eq:legendre_assoc}. 
The phase is chosen as:
\begin{equation}
    \mathcal{P}_\ell^{-|m|}(\theta) = (-1)^m\; \mathcal{P}_\ell^{|m|}(\theta)
\end{equation}
In the case $m=0$, we find the usual reduction to Legendre polynomials:
\begin{eqnarray}
    Y_{\ell,0} (\theta,\,\phi) & = &(2\pi)^{-1/2}\mathcal{P}_\ell^0(\theta) \nonumber \\
    & = & \left( \frac{2\ell+1}{4\pi}  \right)^{1/2} P_\ell(\cos\theta)
    \end{eqnarray}
    
These definitions are used in the normalizations and fit formulas of the \YUMI\, code.

\subsection{Normalization of cross sections}\label{sec:norm_cross}
The normalization of the cross-sections/rates is a delicate matter, since it hinges on the chosen normalization of the associated Lagrange polynomials as well as those of the 6-j Wigner coefficients with respect to the 3-j and Clebsch-Gordon coefficients. Also, because of the statistical nature of the pressure broadening observable \citep{ben-reuven_impact_1966}, one must be aware of the averaging/summing to be used for the quantum measurement.  In our case, with the definition of all special functions and recoupling coefficients used, the normalization for PB sections simplifies and reads:
\begin{equation}\label{eq:app_norm}
    N=\frac{1}{2j'_2+1} ,
\end{equation}
with $j'_2$, the incoming projectile quantum number, if relevant (for He, it would be $j'_2=0$). Note that there is no factor associated with vibrational motion, for it has no degeneracy here. A more elaborate treatment is in order for all more involved cases (like symmetric/asymmetric/spherical rotors or degenerate vibrational modes).

\section{Para- and Ortho-\ce{H2}}\label{sec:ratio}
\begin{figure}[h!]
    \centering
 \includegraphics[width=0.75\linewidth]{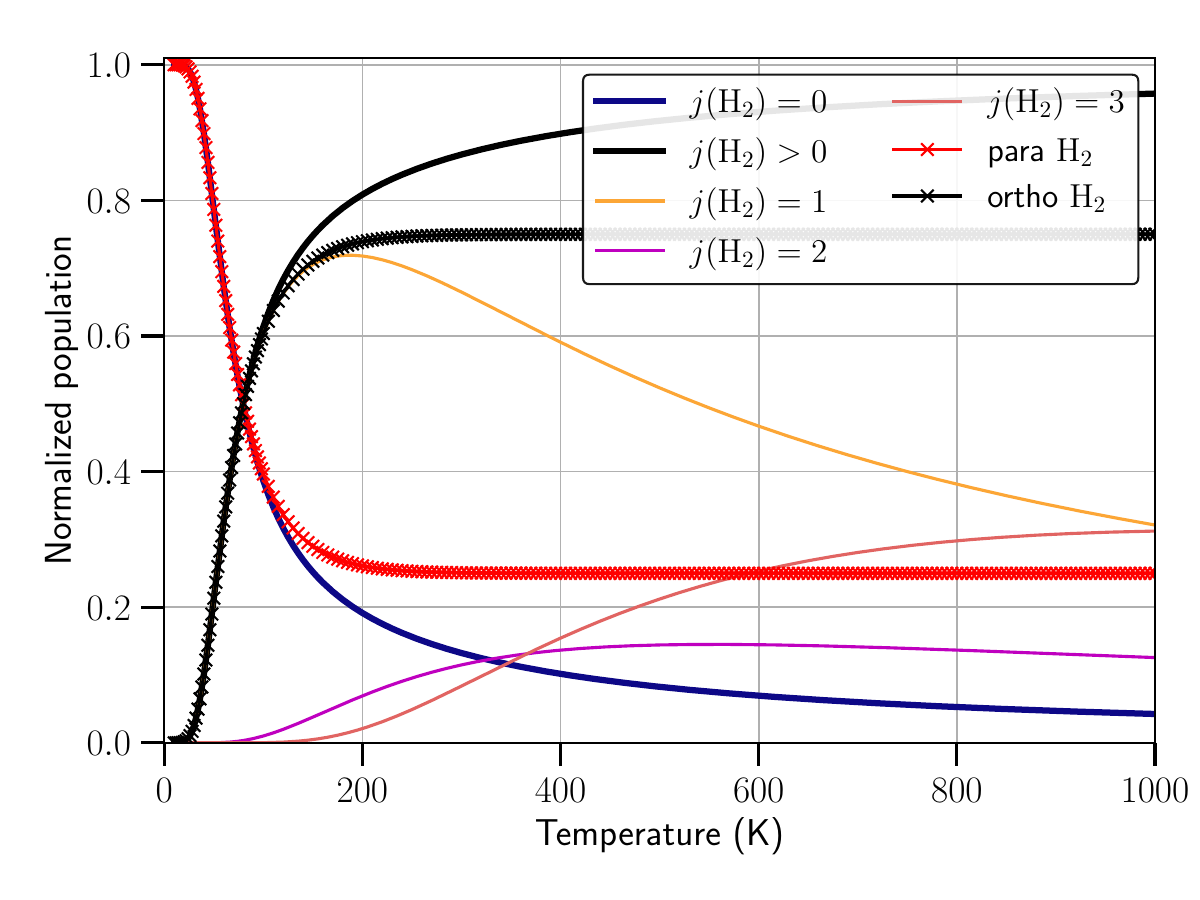}
    \caption{Temperature dependence of the normalized rotational-state populations of molecular hydrogen \ce{H2}. The individual populations for $j$(\ce{H2}) = 0,1,2, and 3 are shown, along with the total para-\ce{H2} population (even $j$) and ortho-\ce{H2} population (odd $j$). At low  ($T\lesssim 80$~K), the $j(\ce{H2})=0$ para state dominates, while increasing temperature progressively populates higher rotational levels and drives the ortho-to-para ratio  towards its high-temperature statistical equilibrium value, $3/1$. Also, at $T>\gtrsim 350$ K, the $j(\ce{H2})>0$ makes up more than 90\% of the \ce{H2} population.}
    \label{fig:Ortho2ParaRatio}
\end{figure}

The pressure broadening coefficients of CO$_2$ due to H$_2$ collisions can be separated into the effects para-\ce{H2} ($j_2=0$) and all the others ($j_2 > 0$) cases, where the latter is well approximated by the ortho-\ce{H2} ($j_2=1$) case. Accordingly, when combining the data, we group the results into ortho- and para- contributions using the ratios presented in \autoref{fig:Ortho2ParaRatio}. At room temperature, the ortho-\ce{H2} fraction is approximately $\sim0.866$, while the remaining $\sim0.134$ corresponds to para-\ce{H2}. Since states with $j_{\ce{H2}} > 0$ increasingly dominate the population at higher temperatures, their pressure broadening coefficients tend to remain much closer to the ortho-\ce{H2} results across the temperature range considered here. The difference between the ortho and para cases remains within the typical $\sim$10\% uncertainty requirement proposed for JWST observations \citep{Wiesenfeld:2025aa}. It must be noted that this way we slightly \emph{under-estimate} the actual coefficients (by a few \%), as the collisions with $j_2\geq 2$ tend to be slightly stronger \cite{thibault_line_2025}, also observed in normal inelastic scattering at high energies \cite{dubernet_rotational_2009,wiesenfeld2022quenching}.

\section{RPA vs Full Calculation}\label{sec:RPA}
\begin{figure}[h!]
    \centering
    \includegraphics[width=\linewidth]{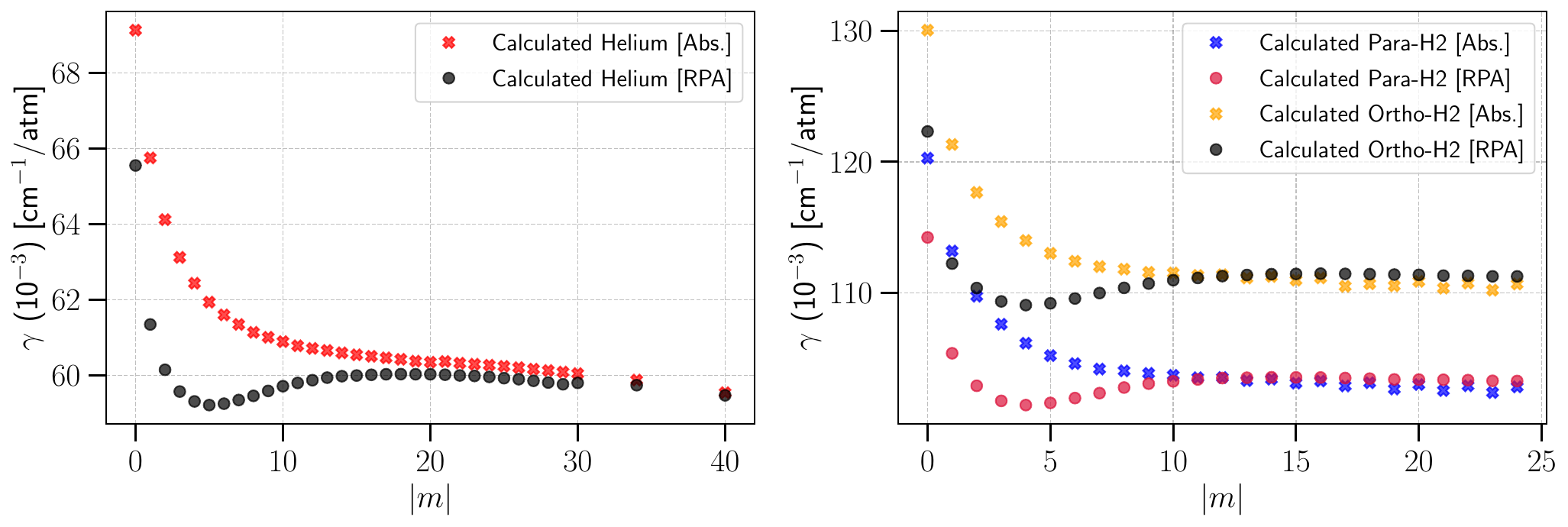}
    \caption{Comparison Random Phase Approximation calculation from \autoref{eq:RPA} to those of the full calculations from \autoref{eq:section}. The RPA calculation converges and is within a few percent at large $|m|$. The left panel shows results for He broadening, while the right panel presents the corresponding results for para-H$_2$ and ortho-H$_2$. }
    \label{fig:RPA}
\end{figure}

For all pressure broadening, the full calculations \autoref{eq:section} exhibit a strong dependence on $|m|$ at low $|m|$, followed by a gradual convergence toward an approximately constant value at larger $|m|$. The RPA calculations \autoref{eq:RPA} reproduce the large $|m|$ behavior very well (as see in \autoref{fig:RPA}), by converging to within a few percent of the full calculations. However, noticeable discrepancies remain at low $|m|$, where the RPA tends to underestimate the broadening coefficients, particularly for He and para-H$_2$.  The agreement improves significantly beyond $|m| \gtrsim 10$, indicating that the RPA captures the dominant high-$|m|$ behavior despite missing some of the stronger coupling effects present in the exact calculations at small $|m|$. This is to be put in line with the high total energy/temperature  behavior of the \ce{H2O-H2} \cite{faure2013pressure} or \ce{CO-H2} \cite{thibault_line_2025}.

\section{Scalability}
\label{sec:scalability}
\begin{figure}[ht!]
    \centering
    \includegraphics[width=0.75\linewidth]{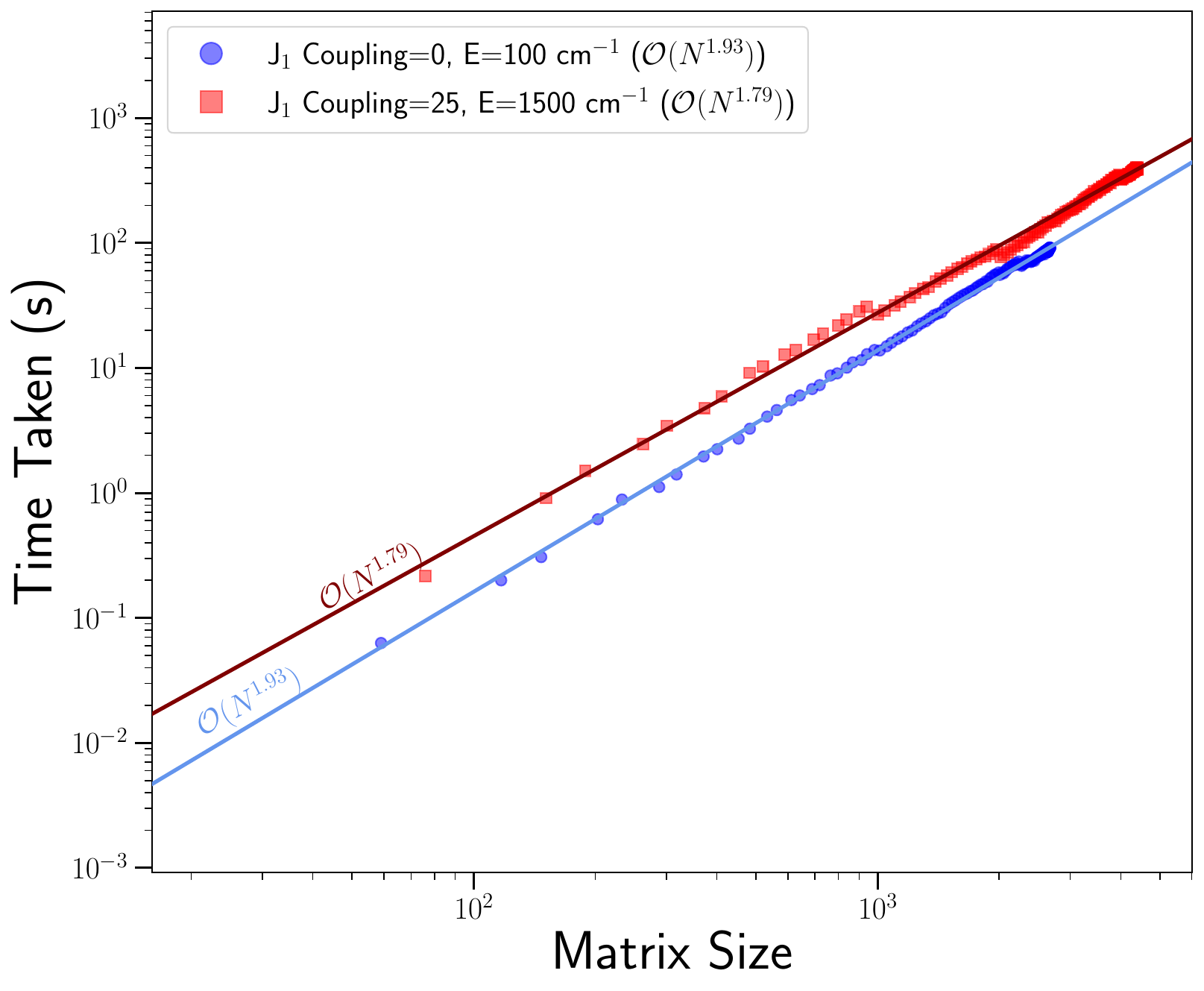}
    \caption{The time taken across different matrix sizes for two computational cases of \COTwo-\HTwo\, calculations at different $J_1$-coupling and collisional energy. The calculation scales roughly at $\mathcal{O}$($\mathcal{N}^2$), with obvious discontinuities occurring near steps of 500,  due to recursive matrix inversion which divides large matrices into sizes of 500 or smaller.}
    \label{fig:scalability}
\end{figure}

The primary computational bottleneck in \YUMI\, arises during the matrix inversion required to solve the time-independent Schr\"odinger's equation. While the detailed implementation will be presented in a dedicated paper on \YUMI\, (Jaïdane et al., in prep), the code uses highly optimized inversion routines from the \texttt{LAPACK} library, built with the recursive algorithm prescribed in \citep{IngemarssonGustafsson2015}. We are able to obtain roughly a scaling factor of $\mathcal{O}$$(\mathcal{N})^2)$ (see \autoref{fig:scalability}). At higher energies (corresponding to higher temperatures and/or higher coupling factors), more energy levels are accessible, which increases both matrix size and computational time. In the current setup, we have successfully computed collisional parameters up to 800 K, representing substantial progress. Additionally, memory usage presents an important consideration. \YUMI\, dynamically allocates the memory for all variables, the largest of which corresponds to the coupling matrix for the potential energy, reaching up to several gigabytes. While the current \COTwo-\HTwo/He~system\ is tractable, future applications involving more complex systems will necessitate stronger consideration for memory management as well as computational load. To address these challenges, we are actively exploring GPU-accelerated algorithms and machine-learning surrogate models to predict pressure broadening parameters for effectively modeling more complex collisional systems.

\section{Tables}\label{sec:tables}
Full results are given in \href{https://doi.org/10.5281/zenodo.20435057}{Zenodo\footnote{\url{https://doi.org/10.5281/zenodo.20435057}}}, where a ReadMe contains the file descriptions. We also add the following for completeness:

\begin{description}
    \item[Files 1] The coefficients of the $V(\ce{CO2-H2})$ PES, in the following format: \\
    Line 1 :  $n $ \newline
    Line 2 :  the $n$ radial distances $r_1,\ldots r_n$\newline
    Following lines : $\ell_1 \> \ell_2 \> \ell $, \{the $n$ coefficients , one for each distance given in line 2 \}.
    \item[Files 2] The coefficients of the $V(\ce{CO2-He})$ PES, in the same format, with $\ell_2\equiv 0$ and $\ell = \ell_1$.
    
    \item[File 3 \& 4] The power coefficients of the pressure for single power law and for double power law from \autoref{tab:gamma}.

    \item[File 5, 6 \& 7] The pressure broadening $\gamma$ values (in $10^{-3}$ \wn/atmosphere) for \ce{CO2-He} \& \ce{CO2-H2} (ortho and para) separately over the range of 40 K to 800 K.
    
    \item[File 8] The RPA calculation of the pressure broadening for all three cases: Case 1 (\ce{CO2}-He), Case 2 (\ce{CO2}-para \ce{H2}) and Case 3 (\ce{CO2}-ortho \ce{H2}) over the range of 40 K to 800 K. 
    
    \item[File 9 \& 10] The rate coefficients of pressure broadening for all three cases: Case 1 (\ce{CO2}-He), Case 2 (\ce{CO2}-para \ce{H2}) and Case 3 (\ce{CO2}-ortho \ce{H2}) over the range of 40 K to 800 K. The ElasticRateCoefficients.txt are for the elastic rate coefficients, while the InelasticRateCoefficients.txt are for the de-excitation.
\end{description}

\bibliography{./Bibfiles/bibliography.bib,./Bibfiles/perso_scholar.bib}

\end{document}